\documentclass[article,shortnames]{jss}
\pdfoutput=1
\usepackage[utf8]{inputenc} 
\usepackage[intlimits, fleqn]{amsmath} 
\usepackage{subcaption} 
\usepackage{graphicx} 
\usepackage[section]{placeins} 
\usepackage[english]{babel}
\usepackage{multirow} 

\author{Robert Geitner\\Friedrich Schiller University Jena \And 
        Robby Fritzsch\\Friedrich Schiller University Jena \AND
        J\"urgen Popp\\Friedrich Schiller University Jena \And
        Thomas W. Bocklitz\\Friedrich Schiller University Jena}
        
\title{\pkg{corr2D} - Implementation of Two-Dimensional Correlation Analysis in \proglang{R}}

\Plainauthor{Robert Geitner, Robby Fritzsch, J\"urgen Popp, Thomas W. Bocklitz}
\Plaintitle{corr2D - Implementation of Two-Dimensional Correlation Analysis in R} 

\Abstract{
In the package \pkg{corr2D} two-dimensional correlation analysis is implemented in \proglang{R}. This paper describes how two-dimensional correlation analysis is done in the package and how the mathematical equations are translated into \proglang{R} code. The paper features a simple tutorial with executable code for beginners, insight into at the calculations done before the correlation analysis, a detailed look at the parallelization of the fast Fourier transformation based correlation analysis and a speed test of the calculation. The package \pkg{corr2D} offers the possibility to preprocess, correlate and postprocess spectroscopic data using exclusively the \proglang{R} language. Thus, \pkg{corr2D} is a welcome addition to the toolbox of spectroscopists and makes two-dimensional correlation analysis more accessible and transparent.
}
\Plainkeywords{correlation analysis, 2D correlation, spectroscopy, R, R package, corr2D}

\Address{
  Robert Geitner\\
  Institute for Physical Chemistry and Abbe Center of Photonics\\
  Faculty of Chemistry and Geography\\
  Friedrich Schiller University Jena\\
  Helmholtzweg 4, 07743 Jena, Germany\\
  \\
  Robby Fritzsch\\
  Institute for Physical Chemistry and Abbe Center of Photonics\\
  Faculty of Chemistry and Geography\\
  Friedrich Schiller University Jena\\
  Helmholtzweg 4, 07743 Jena, Germany\\
  \\
  J\"urgen Popp\\
  Institute for Physical Chemistry and Abbe Center of Photonics\\
  Faculty of Chemistry and Geography\\
  Friedrich Schiller University Jena\\
  Helmholtzweg 4, 07743 Jena, Germany\\
  \\
  Leibniz Institute of Photonic Technology (IPHT) Jena\\
  Albert-Einstein-Str. 9, 07745 Jena, Germany\\
  \\
  Thomas W. Bocklitz\\
  Institute for Physical Chemistry and Abbe Center of Photonics\\
  Faculty of Chemistry and Geography\\
  Friedrich Schiller University Jena\\
  Helmholtzweg 4, 07743 Jena, Germany\\
  \\
  Leibniz Institute of Photonic Technology (IPHT) Jena\\
  Albert-Einstein-Str. 9, 07745 Jena, Germany\\
  E-mail: \email{thomas.bocklitz@uni-jena.de}\\
}

\begin{document}

\section{Introduction to 2D correlation spectroscopy}
Since their invention scientist used infrared (IR), Raman or nuclear magnetic resonance (NMR) spectroscopy to gain information on atoms and molecules. The usual way to extract information from IR, Raman or NMR spectra is to assign observed spectral signals to molecular structures and thus deducing molecular properties. When analyzing a series of spectra it is sometimes difficult to identify spectral changes of two overlapping signals making it impossible to assign these signals to specific molecular structures. To overcome these problems two-dimensional (2D) correlation analysis was invented \citep{Noda.1989,Noda.1993}.\\
The basic idea of a correlation analysis is to analyze how similar (or dissimilar) two spectral signals change. The correlation analysis describes in a quantitative manner how similar these two signals behave. 2D correlation spectroscopy is a pure mathematical processing of signals. 2D NMR or 2D IR experiments which are based on physical correlation processes during the respective spectroscopic measurements are related to 2D correlation spectroscopy. 2D correlation spectroscopy correlates spectroscopic data after the measurement while 2D NMR and 2D IR experiments generate the correlation during the data collection by special experimental setups.\\
2D correlation analysis (which is another term to describe 2D correlation spectroscopy) is used in spectroscopy to analyze spectral features more clearly and to extract additional information, which may be obscured in classical one-dimensional (1D) plots of spectra. To achieve this goal 2D correlation spectroscopy correlates a series of spectra collected under the influence of an external perturbation using the correlation integral. Isao Noda applied the correlation integral to a series of IR spectra of polymers collected under the influence of a sinusoidal tensile strain in 1986 \citep{Noda.1986} and later generalized the approach in 1989 and 1993 \citep{Noda.1989,Noda.1993}.\\
Today 2D correlation analysis is used in spectroscopy to analyze dynamic systems under a specific perturbation. In this context IR, Raman, NMR and UV/vis spectroscopy as well as mass spectrometry have been used to study polymers, reaction solutions and pharmaceuticals under the influence of temperature, time and electro-magnetic radiation. For good reviews on spectroscopic methods, samples and perturbations used in 2D correlation spectroscopy the reader is referred to \cite{Noda.2014, Noda.2014b} and \cite{Park.2016}.\\
Although 2D correlation spectroscopy is used by an ever growing community, there has been to the best of our knowledge no publically available implementation of 2D correlation spectroscopy in \proglang{R} \citep{RCoreteam.2016}. Furthermore there is only one standalone software available to do 2D correlation spectroscopy. It is called 2DShige, can be downloaded for free and was developed by Shigeaki Morita \citep{Morita.2005}. Unfortunately, 2DShige is a standalone program and thus it is difficult to use it in combination with other software, which may be used to preprocess the spectroscopic data accordingly. It is also not an open source software and thus lacks transparency. As an alternative to 2DShige home-written MATLAB \citep{MATLAB.} scripts are often used to carry out 2D correlation analysis \citep{LopezDiez.2005,Barton.2006,Spegazzini.2012} (see also MATLAB contribution MIDAS 2010 by \cite{Ferenc.2011}). These MATLAB scripts allow the user to preprocess and correlate the data within one program, but also lack transparency and comprehensibility. The spectroscopy and analysis software OPUS from \cite{BrukerCorporation.} also has an implemented 2D correlation spectroscopy algorithm. Unfortunately, OPUS is a commercial software and lacks some freedom as well as transperancy, which other statistical software like \proglang{R} or MATLAB offer. Thus, OPUS is very rarely used to perform 2D correlation spectroscopy. The widespread analysis software Origin by \cite{OriginLab.} offers some related correlation analysis, but lack (up to 2016) the implementation of 2D correlation spectroscopy as described by Noda.\\
In this paper we present our \proglang{R}~package \pkg{corr2D} \citep{Geitner.2016b}, which implements 2D correlation spectroscopy in \proglang{R} and is available on the Comprehensive \proglang{R} Archive Network (CRAN). Thus the package \pkg{corr2D} combines transparency, comprehensibility and the convenience to process and analyze 2D correlation spectra within one open source program. We already published some results \citep{Geitner.2015, Geitner.2016} utilizing the calculation and plotting properties of \pkg{corr2D}. For the calculation of the complex correlation matrix a parallelized fast Fourier transformation (FFT) approach is used. To illustrate the use of \pkg{corr2D} the package also features a set of preprocessed temperature-dependent experimental Raman spectra \citep{Geitner.2015} and a function to generate artificial data. We hope to enrich both the \proglang{R}~community as well as the 2D correlation community with our package.\\
The paper is divided into three main sections. The first section deals with the mathematical background and the theoretical description of 2D correlation spectroscopy. The comphrensive mathematical description of 2D correlation spectroscopy is important because the package \pkg{corr2D} translates the 2D correlation theory into executable \proglang{R}~code. For newcomers to the field of 2D correlation analysis we suggest reading of \cite{Noda.2000b} as it features a simplified introduction to the formal mathematical procedure and three application examples. The second section is meant as a tutorial for beginners. It describes the structure of the input data and how the resulting object containing the 2D correlation spectra can be visualized. In addition the arguments of the plotting function \code{plot\_corr2d()} and \code{plot\_corr2din3d()} are presented. To round out the tutorial the section also gives a short introduction to the interpretation of 2D correlation spectra. The third and final section of the paper further dives into the technical details of \pkg{corr2D}. The section focuses on how the mathematical equations described in the first section are translated into \proglang{R}~code, how special features of 2D correlation spectroscopy are implemented into \code{corr2d()}, how the 2D correlation analysis was parallelized and how fast the resulting \proglang{R}~code is. The final section also explains the \proglang{R}~code behind the plotting function \code{plot\_corr2d()} and \code{plot\_corr2din3d()}.\\ 

\section{Theoretical description of 2D correlation spectroscopy}
\label{sec:theo}
The foundation of 2D correlation spectroscopy are the general auto- and cross-correlation integrals seen in Equations \ref{eq:autocorr} and \ref{eq:crosscorr}. The result of a general correlation analysis is the correlation coefficient $C$ which describes how similar two signals $f(u)$ and $g(u)$ are depending on a lag $\tau$ between them. $f^*(u)$ denotes the complex conjugate of $f(u)$.

\begin{equation} \label{eq:autocorr} C_{\mathrm{auto}}(\tau) = \displaystyle\int_{-\infty}^\infty f^*(u) \cdot f(u+\tau)du \end{equation}
\begin{equation} \label{eq:crosscorr} C_{\mathrm{cross}}(\tau) = \displaystyle\int_{-\infty}^\infty f^*(u) \cdot g(u+\tau)du \end{equation}

To use the general correlation integral on spectroscopic data the integral needs to be specified. The idea is that the dynamic variations of two signals $y_1(\nu_1, t)$ and $y_2(\nu_2, t)$ (inserted as $f(u)$ and $g(u)$ in Equation \ref{eq:crosscorr}) both examples of spectra depending on their own spectral variables $\nu_1$ and $\nu_2$ as well as on an external perturbation variable $t$ should be correlated with each other. The spectra are observed within a perturbation interval ranging from $T_{\mathrm{min}}$ to $T_{\mathrm{max}}$. This interval is used together with the reference spectrum $\overline{y}(\nu)$ to formally define the dynamic spectrum $\widetilde{y}(\nu, t)$ as seen in Equation \ref{eq:dynspec}. The dynamic spectra represent the dynamic changes observed within the signals $y_1(\nu_1, t)$ and $y_2(\nu_2, t)$ induced by the perturbation $t$.

\begin{equation} \label{eq:dynspec}
\widetilde{y}(\nu, t) =
  \begin{cases}
    y(\nu, t) - \overline{y}(\nu) & \quad \text{for } T_{\mathrm{min}} \leq t \leq T_{\mathrm{max}}\\
    0                             & \quad \text{otherwise}\\
  \end{cases}
\end{equation}

The reference spectrum $\overline{y}(\nu)$ can be chosen arbitrarily. Often the perturbation mean spectrum is used as the reference spectrum (see Equation \ref{eq:ref}). Other reference spectra could be spectra taken before or after the collection of the perturbation dependent spectra series.

\begin{equation} \label{eq:ref} \overline{y}(\nu) = \frac{1}{T_{\mathrm{max}}-T_{\mathrm{min}}} \displaystyle\int_{T_{\mathrm{min}}}^{T_{\mathrm{max}}} y(\nu, t) dt\end{equation}

There are in principle two ways to calculate 2D correlation spectra: The first approach is based on the Fourier transformation (FT) \citep{Noda.1993}, while the second one uses the Hilbert transformation (HT) \citep{Noda.2000b}. The results of both approaches are identical.\\
Following the FT approach the dynamic spectra need to be Fourier transformed to separate them into component waves as stated in \cite{Noda.1990} and as can be seen in Equation \ref{eq:FT}. The FTs of the dynamic spectra can then be used to obtain a complex cross correlation function (Equation \ref{eq:int2d}). The real and imaginary part of the complex cross correlation function are termed synchronous and asynchronous 2D correlation spectra $\Phi(\nu_1, \nu_2)$ and $\Psi(\nu_1, \nu_2)$.

\begin{equation} \label{eq:FT} \widetilde{Y}(\nu, \omega) = \mathcal{F}(\widetilde{y}(\nu, t)) = \displaystyle\int_{-\infty}^\infty \widetilde{y}(\nu, t) \cdot e^{-i\omega t} dt \end{equation}
\begin{equation} \label{eq:int2d} \Phi(\nu_1, \nu_2) + i\Psi(\nu_1, \nu_2) = \frac{1}{2 \pi (T_{\mathrm{max}} - T_{\mathrm{min}})} \displaystyle\int_{-\infty}^\infty \widetilde{Y}(\nu_1, \omega) \cdot \widetilde{Y}^*(\nu_2, \omega) d\omega \end{equation}

\begin{figure}[tp]
\centering
\includegraphics[width = 1\textwidth]{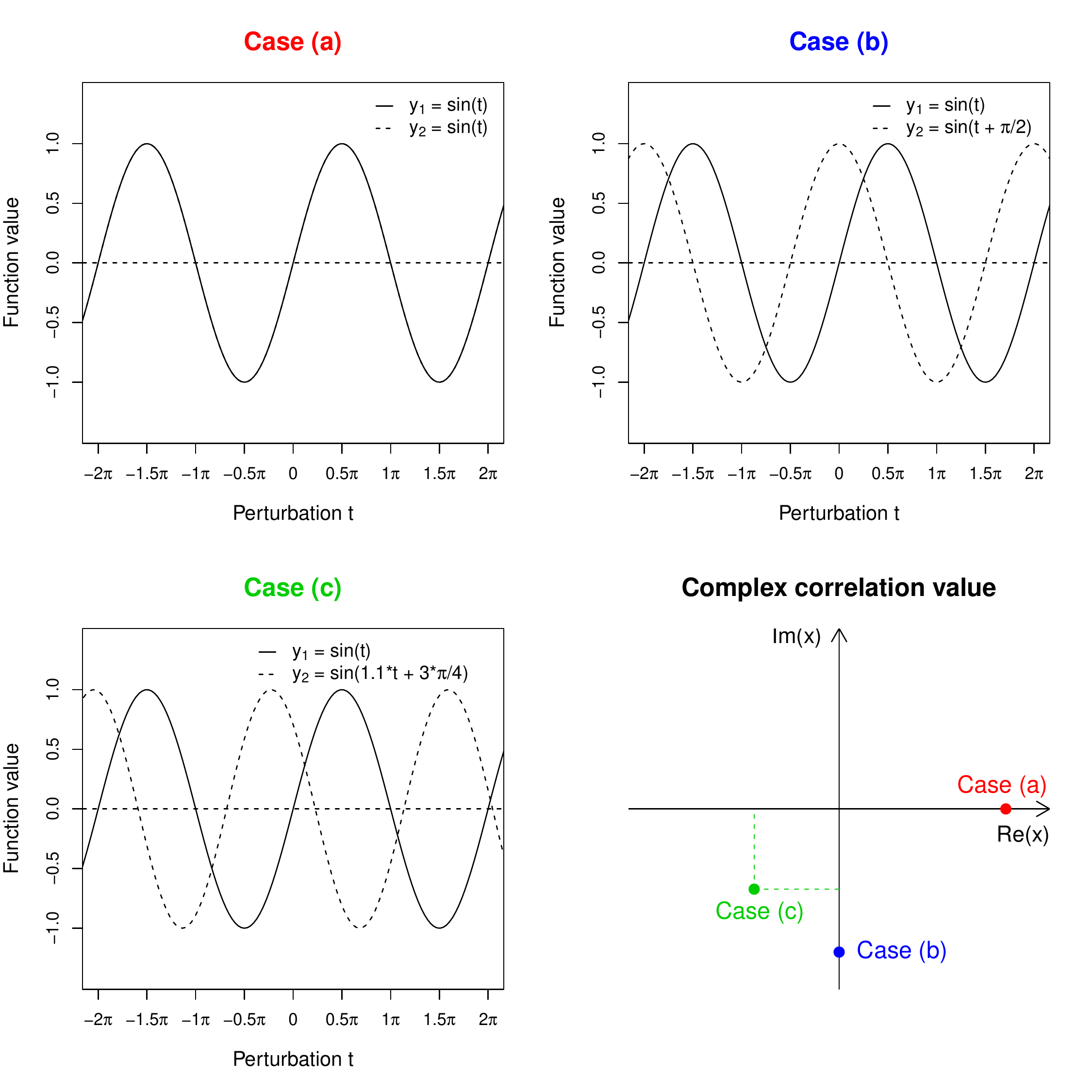}
\caption{The figure illustrates three examples of a correlation analysis of two signals using Equations \ref{eq:disFT} and \ref{eq:disint2d}. The signals $y_1$ and $y_2$ react to a pertubation $t$. Case (a) (red; top left panel) shows a pure synchronous behavior, while case (b) (blue; top right panel) illustrates the pure asynchronous behavior. Case (c) (green; bottom left panel) showcases the ordinary correlation behavior where the complex correlation value (bottom right panel) shows synchronous and asynchronous contributions. For details see the text.}
\label{fig:correxam}
\end{figure}

Figure \ref{fig:correxam} illustrates how two signals $y_1(t)$ and $y_2(t)$ both depending a perturbation $t$ are correlated with each other. Both signals react to the external perturbation. This could be two specfic wavenumber positions in a Raman spectrum reacting to a changing temperature. If both signals react identical to the perturbation (case (a) in Figure \ref{fig:correxam}) the resulting complex correlation value has a non-zero real part while the imaginary part is zero. According to Equation \ref{eq:int2d} the real and imaginary part of the complex correlation value are called synchronous and asynchronous 2D correlation intensities $\Phi$ and $\Psi$. If both signals react exactly with a phase difference of $\pi/2$ to the perturbation (case (b)) then the complex correlation value only consits of an imaginary part. This means that the two signals only show an asynchronous correlation behavior. The case that is most often encountered when analyzing real-world data is that the complex correlation coefficient is made up from both real and imaginary parts and that the two correlated signals show synchronous as well as asynchronous correlation behavior (case (c) in Figure \ref{fig:correxam}). During the process of a complete 2D correlation analysis not only two but all combinations of spectral signals are correlated with each other. To make the results accessible to humans the real and imaginary parts of the calculated complex correlation coefficients are presented as synchronous and asynchronous 2D correlation spectra.

When analyzing $m$ discrete data values the FT has to change to the discrete Fourier transformation (DFT). Following this change Equations \ref{eq:FT} and \ref{eq:int2d} transform into Equations \ref{eq:disFT} and \ref{eq:disint2d}. A fast implementation of the DFT is the fast Fourier transformation (FFT), which is often used to implement the DFT within computer algorithms. When using Equation \ref{eq:disint2d} for the calculation of 2D correlation spectra of discrete data an important condition to be fulfilled is the even spacing of the discrete perturbation values $T$ along the perturbation axis $t$. Otherwise the unevenly sampled data has to be interpolated to form evenly sampled data or the correlation function needs to be modified \citep{Noda.2003}.

\begin{equation} \label{eq:disFT} \widetilde{Y}(\nu, \omega) = \displaystyle\sum_{j = 1}^{m} \widetilde{y}(\nu, t_j) \cdot e^{-i \frac{2 \pi \cdot \omega}{m} (j - 1)} \quad \text{for } \omega = 0, 1, \dots, m-1 \end{equation}
\begin{equation} \label{eq:disint2d} \Phi(\nu_1, \nu_2) + i\Psi(\nu_1, \nu_2) = \frac{1}{\pi(m - 1)} \displaystyle\sum_{\omega = 0}^{m-1} \widetilde{Y}(\nu_1, \omega) \cdot \widetilde{Y}^*(\nu_2, \omega) \end{equation}

If the input data is real (a condition that is often met when dealing with real world data) the DFT features Hermitian symmetry as seen in Equation \ref{eq:symmeFT}, which can be used to further simplify the calculation process.

\begin{equation} \label{eq:symmeFT} \widetilde{Y}(\nu, \omega) = \widetilde{Y}^*(\nu, m - \omega) \quad \text{for } \omega = 1, 2, \dots, m - 1 \end{equation}

In summary the calculation of 2D correlation spectra following the FT approach consists of three main steps:

\begin{enumerate}
\item Calculation of the dynamic spectra $\widetilde{y}(\nu, t)$ from the preprocessed spectra $y(\nu, t)$ and a chosen reference spectrum $\overline{y}(\nu)$.
\item FT of the dynamic spectra $\widetilde{y}(\nu, t)$ to receive $\widetilde{Y}(\nu, \omega)$.
\item Correlation of the Fourier transformed dynamic spectra $\widetilde{Y}(\nu, \omega)$ for every spectral value pair using the correlation integral to calculate the synchronous and asynchronous correlation spectra $\Phi(\nu_1, \nu_2)$ and $\Psi(\nu_1, \nu_2)$.
\end{enumerate}

Another approach to calculate 2D correlation spectra is based on the HT \citep{Noda.2000b}. The HT approach splits the calculation of the synchronous and asynchronous 2D correlation spectra. The synchronous 2D correlation spectrum $\Phi(\nu_1, \nu_2)$ can be directly calculated from the dynamic spectra (see Equation \ref{eq:HTsyn}). For the calculation of the asynchronous 2D correlation spectrum $\Psi(\nu_1, \nu_2)$ the HT $\widetilde{z}(\nu_2, t)$ of one of the dynamic spectra is needed (see Equations \ref{eq:HT} and \ref{eq:HTasyn}). The HT is defined using the Cauchy principle value.

\begin{equation} \label{eq:HTsyn} \Phi(\nu_1, \nu_2) = \frac{1}{T_{\mathrm{max}} - T_{\mathrm{min}}} \displaystyle\int_{T_{\mathrm{min}}}^{T_{\mathrm{max}}} \widetilde{y}(\nu_1, t) \cdot \widetilde{y}(\nu_2, t) dt \end{equation}
\begin{equation} \label{eq:HT} \widetilde{z}(\nu_2, t) = \mathcal{H}(\widetilde{y}(\nu_2, t)) = \frac{1}{\pi} \quad PV\displaystyle\int_{-\infty}^\infty \frac{\widetilde{y}(\nu_2, t')}{t'-t} dt \end{equation}
\begin{equation} \label{eq:HTasyn} \Psi(\nu_1, \nu_2) = \frac{1}{T_{\mathrm{max}}-T_{\mathrm{min}}} \displaystyle\int_{T_{\mathrm{min}}}^{T_{\mathrm{max}}} \widetilde{y}(\nu_1, t) \cdot \widetilde{z}(\nu_2, t) dt \end{equation}

When dealing with $m$ discrete data values Equations \ref{eq:HTsyn}~-~\ref{eq:HTasyn} can be transformed to their respective discrete forms as seen in Equations \ref{eq:disHTsyn}~-~\ref{eq:disHTasyn}. The discrete HT can be done using the so called Hilbert-Noda matrix $N_{jk}$. The Hilbert-Noda matrix simplifies the discrete HT to a matrix vector multiplication. As discussed for the FT approach it is also important for the HT approach that the discrete data is equidistant \citep{Noda.2003}.

\begin{equation} \label{eq:disHTsyn} \Phi(\nu_1, \nu_2) = \frac{1}{m - 1} \displaystyle\sum_{j = 1}^{m} \widetilde{y}(\nu_1, t_j) \cdot \widetilde{y}(\nu_2, t_j) \end{equation}
\begin{equation} \label{eq:disHT} \widetilde{z}(\nu_2, t_j) = \displaystyle\sum_{k = 1}^{m} N_{jk} \cdot \widetilde{y}(\nu_2, t_k) \end{equation}
\begin{equation} \label{eq:HilbNoda}
N_{jk} =
  \begin{cases}
    0                     & \quad \text{if } j = k\\
    \frac{1}{\pi (k-j)}   & \quad \text{otherwise}\\
  \end{cases}
\end{equation}
\begin{equation} \label{eq:disHTasyn} \Psi(\nu_1, \nu_2) = \frac{1}{m - 1} \displaystyle\sum_{j = 1}^{m} \widetilde{y}(\nu_1, t_j) \cdot \widetilde{z}(\nu_2, t_j) \end{equation}

In summary the calculation of 2D correlation spectra following the HT approach consists of four main steps:

\begin{enumerate}
\item Calculation of the dynamic spectra $\widetilde{y}(\nu, t)$ from the preprocessed spectra $y(\nu,t)$ and the reference spectrum $\overline{y}(\nu)$.
\item Calculation of the synchronous 2D correlation spectrum $\Phi(\nu_1, \nu_2)$ by multiplying $\widetilde{y}(\nu_1, t)$ and $\widetilde{y}(\nu_2, t)$.
\item HT of the dynamic spectra $\widetilde{y}(\nu_2, t)$ to get $\widetilde{z}(\nu_2, t)$.
\item Calculation of the asynchronous 2D correlation spectrum $\Psi(\nu_1, \nu_2)$ by multiplying $\widetilde{y}(\nu_1, t)$ and $\widetilde{z}(\nu_2, t)$.
\end{enumerate}

As stated above the result of the FT and HT approach are identical. Thus the 2D correlation spectra display the same information independent of how they were calculated. The synchronous 2D correlation spectrum $\Phi(\nu_1, \nu_2)$ shows which spectral features change in-phase while the asynchronous 2D correlation spectrum $\Psi(\nu_1, \nu_2)$ shows which spectral features change out-of-phase. If a spectral dataset is correlated with itself (an autocorrelation in the general terminology) the resulting 2D spectra are called homo correlation spectra. If two different spectral datasets are correlated with each other (a cross correlation in the general terminology) the resulting 2D spectra are called hetero correlation spectra.\\

In addition to the basic 2D correlation spectroscopy described above, modifications to the original technique have been developed. One of these modifications is the application of scaling techniques to 2D correlation spectra. The goal of scaling 2D correlation spectra is to enhance correlation signals with low intensity when compared to correlation signals with high intensity. High intensity signals sometimes dominate 2D correlation spectra and thus hamper the identification of smaller signals. The approach to solve this problem is described in \cite{Noda.2008}. The basic idea is to scale the synchronous and asynchronous correlation spectra $\Phi(\nu_1, \nu_2)$ and $\Psi(\nu_1, \nu_2)$ using the standard deviation $\sigma_{\nu}$, which is calculated from the original spectral dataset $y(\nu)$ and the reference spectrum $\overline{y}(\nu)$ as seen in Equation \ref{eq:stdev}. Strictly this scaling is only defined when using the mean-spectrum (see Equation~\ref{eq:ref}) as reference spectrum, but the scaling procedure can also applied when using other reference spectra. Care should be taken when interpreting scaled 2D correlation spectra which are not scaled using the perturbation mean-spectrum.

\begin{equation} \label{eq:stdev} \sigma_{\nu_i} = \sqrt{\frac{\displaystyle\sum_{j=1}^m(y(\nu_i, t_j)-\overline{y}(\nu))^2 }{m - 1}} \end{equation}

For every spectral value $\nu_i$ (for $i = 1, 2, \dots, n$) there is a standard deviation $\sigma_{\nu_i}$. To scale the correlation intensity at the position $(\nu_1, \nu_2)$ the correlation intensity at this position $\Phi(\nu_1, \nu_2)$ (or $\Psi(\nu_1, \nu_2)$) is divided by the product of the two standard deviations $\sigma_{\nu_1}$ and $\sigma_{\nu_2}$ sometimes referred to as total joint variance \citep{Noda.2004}. This conventional unit-variance scaling is called Pearson scaling (see Equations \ref{eq:pearsonsyn} and \ref{eq:pearsonasyn}). Unfortunately, Pearson scaling strongly increases the influence of noise on 2D correlation spectra. To counteract this effect Noda suggested using Pareto scaling where the data is scaled by the square root of the standard deviations. The generalized approach as seen in Equations \ref{eq:scalesyn} and \ref{eq:scaleasyn} is to introduce an exponent $\alpha$ which describes how the total joint variance is used to scale the correlation intensities.

\begin{equation} \label{eq:pearsonsyn} \Phi(\nu_1, \nu_2)^{Pearson} = \frac{\Phi(\nu_1, \nu_2)}{(\sigma_{\nu_1} \cdot \sigma_{\nu_2})^{-1}} \end{equation}
\begin{equation} \label{eq:pearsonasyn} \Psi(\nu_1, \nu_2)^{Pearson} = \frac{\Psi(\nu_1, \nu_2)}{(\sigma_{\nu_1} \cdot \sigma_{\nu_2})^{-1}} \end{equation}
\begin{equation} \label{eq:scalesyn} \Phi(\nu_1, \nu_2)^{(Scaled)} = \frac{\Phi(\nu_1, \nu_2)}{(\sigma_{\nu_1} \cdot \sigma_{\nu_2})^{-\alpha}} \end{equation}
\begin{equation} \label{eq:scaleasyn} \Psi(\nu_1, \nu_2)^{(Scaled)} = \frac{\Psi(\nu_1, \nu_2)}{(\sigma_{\nu_1} \cdot \sigma_{\nu_2})^{-\alpha}} \end{equation}

For more information on the theory of 2D correlation spectroscopy, the calculations, scaling techniques and the influence of noise the reader is referred to the literature \citep{Noda.1990, Noda.2006, Noda.2016, Sasic.2001, Yu.2003, Noda.2008}. For techniques derived from the basic 2D correlation spectroscopy like sample-sample 2D correlation spectroscopy, moving window 2D correlation spectroscopy, multiple-perturbation 2D correlation spectroscopy, double 2D correlation spectroscopy, 2D codistribution spectroscopy and quadrature 2D correlation spectroscopy the reader is also referred to the literature \citep{Sasic.2000, Sasic.2000b, Thomas.2000, Shinzawa.2009, Noda.2010b, Noda.2014, Noda.2016b, Noda.2016c}.\\

\section[corr2D tutorial and interpretation of 2D correlation spectra]{\pkg{corr2D} tutorial and interpretation of 2D correlation spectra}
\label{sec:tutorial}

\sloppy{The package \pkg{corr2D} contains a calculation function \code{corr2d()}, two plotting functions \code{plot\_corr2d()} and \code{plot\_corr2din3d()} as well as two S3 methods \code{summary()} and \code{plot()} for the resulting 2D correlation object.}\\
To get started the user just has to supply an [$m \times n$] matrix containing the data to \code{corr2d()}. The input matrix needs to contain the data as follows: $m$ perturbation values $T$ by rows and $n$ spectral values $\nu$ by columns. The column names of the input matrix should contain the spectral value names. Alternatively, the spectral value names can be specified at the argument \code{Wave1}. The perturbation values can be included as row names. The FuranMale dataset from \pkg{corr2D} can be viewed as an example. It contains temperature-dependent Raman spectra, therefore the perturbation variable $t$ is the temperature and the spectral variable $\nu$ is the relative wavenumber.

\begin{Schunk}
\begin{Sinput}
R> library("corr2D")
R> data("FuranMale")
R> FuranMale[, 1:5]
\end{Sinput}
\begin{Soutput}
       1550.26392    1550.74602   1551.22812   1551.71023  1552.19233
110  0.0058962811  5.506783e-03 0.0051347609 0.0047584418 0.004295668
120  0.0043970716  4.860985e-03 0.0052363316 0.0055872210 0.005806852
130  0.0055330645  4.916008e-03 0.0045517037 0.0043091089 0.004450099
140 -0.0008350893 -4.653592e-04 0.0003397819 0.0014053808 0.002903903
150 -0.0005203668  3.312193e-05 0.0003998590 0.0008810001 0.001332789
160  0.0060360763  6.776913e-03 0.0073772994 0.0076989039 0.007883910
\end{Soutput}
\begin{Sinput}
R> twod <- corr2d(FuranMale)
\end{Sinput}
\begin{Soutput}
HOMO-Correlation: 2 cores used for calculation
10:39:26 - using mean values as reference
10:39:26 - Fast Fourier Transformation and multiplication 
 to obtain a 145 x 145 correlation matrix 
10:39:26 - Done
\end{Soutput}
\end{Schunk}

\begin{figure}[htp]
    \centering
    \begin{subfigure}{0.825\textwidth}
        \centering
        \includegraphics[scale = 1.0]{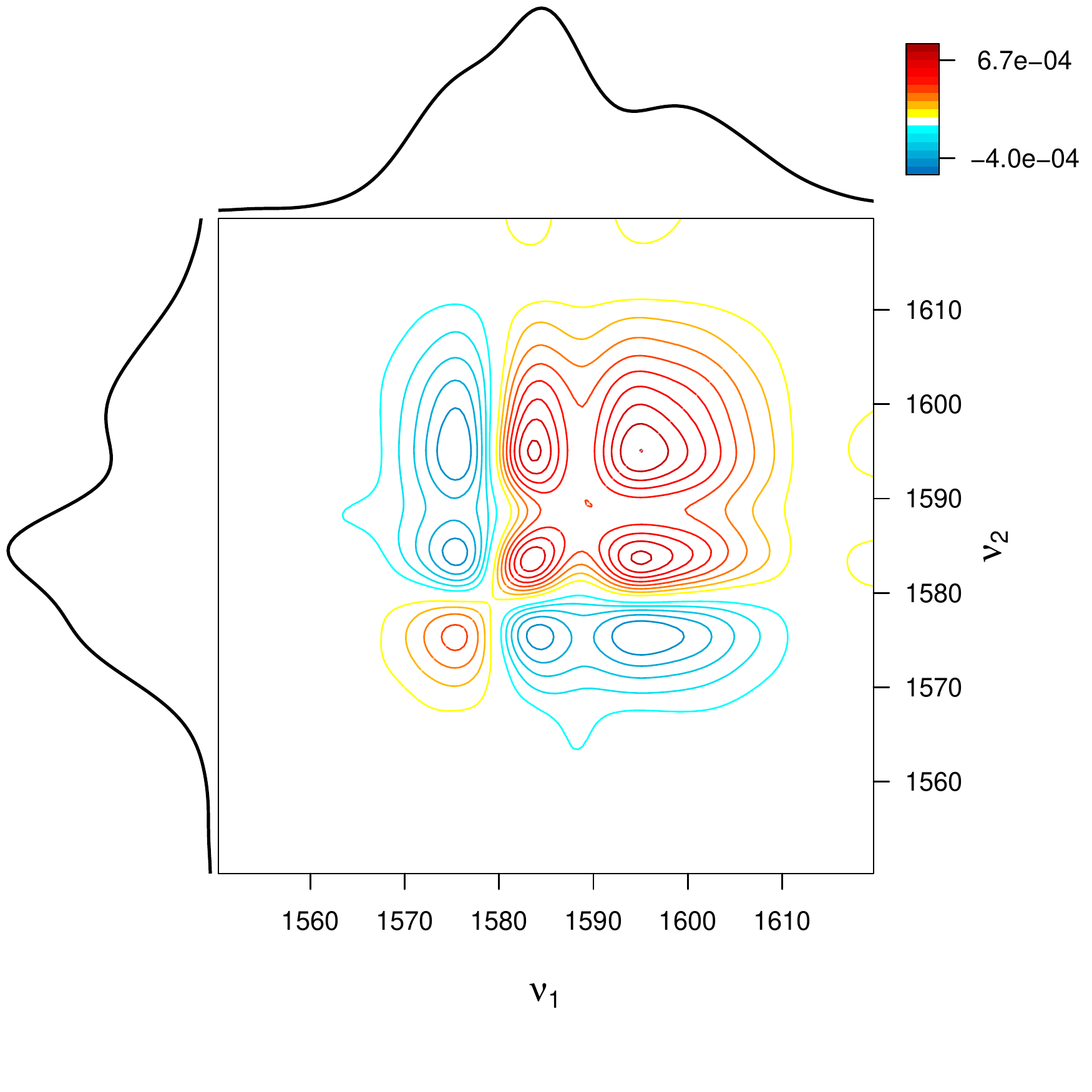}
    \end{subfigure}
    \begin{subfigure}{0.825\textwidth}
        \centering
        \includegraphics[scale = 1.0]{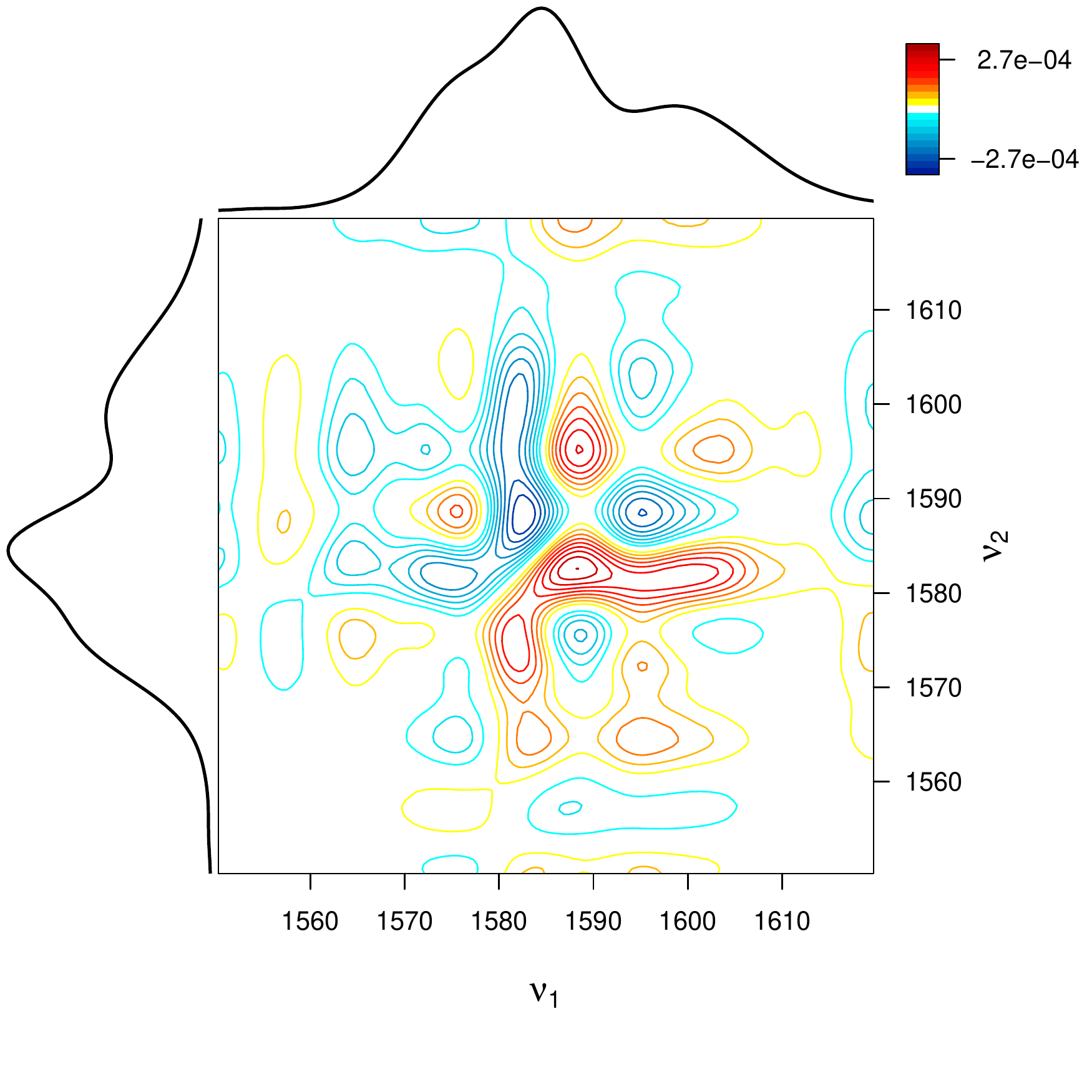}
    \end{subfigure}
    \caption{Synchronous (top) and asynchronous (bottom) homo 2D correlation spectra of the dataset FuranMale. The reference spectrum is the mean spectrum.}
    \label{fig:asynbasic}
\end{figure}

\code{corr2d()} identifies that only one input matrix was given. Thus, \code{corr2d()} correlates the input data with itself which results in homo correlation spectra. If a second input matrix would be given to \code{corr2d()} (at argument \code{Mat2}) the function would automatically calculate the hetero correlation spectra from the two input matrices. Additionally the function also detects that no reference spectrum (at argument \code{Ref1}) was specified and thus builds the mean spectrum, which is then used as reference spectrum. The resulting correlation matrix has the dimensions [$n \times n$], therefore the FuranMale matrix results in a [145 $\times$ 145] correlation matrix. The correlation function \code{corr2d()} will be discussed in detail in Section~\ref{sec:eqtoRcode}.\\
\sloppy{The correlation spectra can be plotted using the \code{plot()} command, which in turn calls \code{plot\_corr2d()} via S3 method dispatch. The real part of the complex correlation spectrum is called synchronous correlation spectrum , while the imaginary part is called asynchronous correlation spectrum (see Figure~\ref{fig:asynbasic}). By default \code{plot\_corr2d()} displays the synchronous spectrum.}

\begin{Schunk}
\begin{Sinput}
R> plot_corr2d(twod)
\end{Sinput}
\end{Schunk}

\begin{Schunk}
\begin{Sinput}
R> plot(twod, Im(twod$FT))
\end{Sinput}
\end{Schunk}

The plot function \code{plot\_corr2d()} offers a lot of features which can be used to alter the appearance of the 2D correlation plot. With the arguments \code{specx} and \code{specy} the data plotted at the top and on the left of the spectrum can be controlled. \code{xlim}, \code{ylim}, \code{zlim}, \code{axes}, \code{xlab} and \code{ylab} are inspired by the normal \code{plot()} (or better \code{image()}) function as they allow the user to control the displayed region as well as the \textit{x}- and \textit{y}-axis and their labels. By default \code{plot\_corr2d()} uses the \code{contour()} function from package \pkg{graphics} to produce a contour plot. This can be changed via the \code{Contour} argument to produce an image (once again using the \pkg{graphics} package) representing the correlation spectrum. The number of equally spaced contour levels (or image layers) plotted can be adjusted with \code{N} and the \code{Cutout} argument. The \code{Cutout} argument can be used to define a number range which will not be plotted. This argument suppresses the plotting of smaller signals and allows to make the resulting plot more clear for the reader, but can also be used to remove unwanted signals and obscure results. Thus, it should always be used with care to make plots more clear without omitting important information. Finally the argument \code{Legend} controls the color legend in the top right corner. Usually the absolute values of correlation spectra are not relevant for the 2D correlation plot. Therefore, the legend can usually be omitted, because the graphical interpretation does not depend on it. For an example with most of the arguments in action see the following code snippet.

\begin{Schunk}
\begin{Sinput}
R> plot(twod, Re(twod$FT), xlim = c(1560, 1620), ylim = c(1560, 1620),
+   xlab = expression(paste("relative Wavenumber" / cm^-1)),
+   ylab = expression(paste("relative Wavenumber" / cm^-1)),
+   Contour = FALSE, N = 32, Cutout = c(-0.8 * 10^-4, 1.3 * 10^-4),
+   Legend = FALSE)
\end{Sinput}
\end{Schunk}

\begin{figure}
\centering
\includegraphics{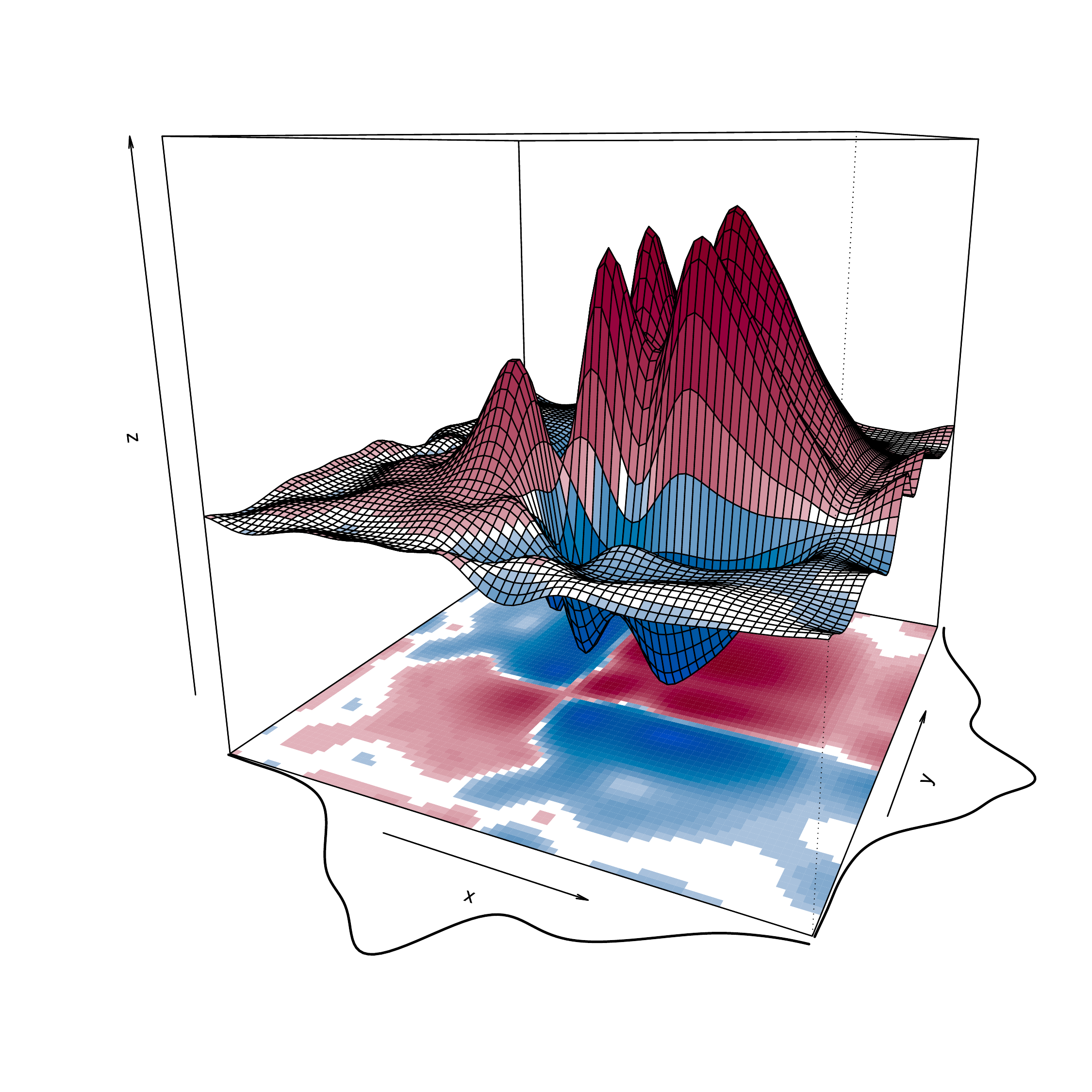}
\caption{3D Surface plot of the synchronous 2D homo correlation spectrum of the dataset FuranMale showcasing \code{plot\_corr2din3d()}. The plot features two custom spectra on the \textit{x}- and \textit{y}-axes and a 2D projection of the 3D surface.}
\label{fig:3dplot}
\end{figure}

Besides \code{plot\_corr2d()} \pkg{corr2D} features another plotting function: \code{plot\_corr2din3d()}. This function can be used to draw a colored 3D surface representation of 2D correlation spectra. To achieve this \code{plot\_corr2din3d()} makes use of \code{drape.plot()} and \code{drape.color()} from package \pkg{fields}. These two functions calculate the color values and graphical positions form the input 2D correlation data and hand these values over to \code{persp()} which than does the plotting of the 3D surface using the perspective angles \code{theta} and \code{phi}. To add information to the plot \code{plot\_corr2din3d()} allows the user to add custom spectra (\code{specx} and \code{specy}) to the \textit{x}- and \textit{y}-axis of the \code{persp()} plot. These spectra can be scaled using the \code{scalex} and \code{scaley} arguments from \code{plot\_corr2din3d()}. The sign of the scaling factor defines if the spectra are plotted inside (positive sign) or outside (negative sign) the \code{persp()} plot. In addition a 2D projection of the 3D surface can be added to the bottom of the plot. This is also done using \code{drape.plot()} and together with the \textit{x}- and \textit{y}-axis spectra the projection recalls the look of a flat 2D correlation plot. To reduce the computational demand of large 2D correlation matrices \code{plot\_corr2din3d()} features the argument \code{reduce} which can be used to reduce the number of points used for drawing the 3D surface. Thus, the argument \code{reduce} allows to calculate a first draft of the 3D surface to adjust the plotting parameters without a high computational demand. Overall, \code{plot\_corr2din3d()} is a useful addition to \pkg{corr2D} and allows to illustrate 2D correlation spectra with impressive colored 3D surface plots. Figure~\ref{fig:3dplot} shows a 3D surface plot of the synchronous 2D correlation spectrum from the FuranMale dataset.

\begin{Schunk}
\begin{Sinput}
R> plot_corr2din3d(Mat = Re(twod$FT), specx = twod$Ref1,
+   specy = twod$Ref1, reduce = 2, scalex = -150, scaley = -130,
+   zlim = c(-0.7, 1) * 10^-3, projection = TRUE,
+   border = NULL, theta = 25, phi = 15, add.legend = FALSE,
+   Col = colorspace::diverge_hcl(129, h = c(240, 0), c = 100,
+     l = c(20, 100), power = 0.3))
\end{Sinput}
\end{Schunk}

The interpretation of 2D correlation spectra is based on the results of the correlation integral. Following the definition of the synchronous and asynchronous correlation spectra (see Figure \ref{fig:correxam} as well as Equations \ref{eq:int2d}, \ref{eq:HTsyn} and \ref{eq:HTasyn}), a set of rules can be derived, which can be used to understand the colorful 2D correlation spectra. These interpretation rules are called 'Noda rules'.\\
There are two types of signals in 2D correlation spectra: Auto peaks and cross peaks. Auto peaks are located at the diagonal of homo 2D correlation, therefore both spectral variables $\nu_1$ and $\nu_2$ have the same value at auto peaks (case (a) in Figure \ref{fig:correxam}). Auto peaks are always non-negative in the synchronous correlation spectrum and are always 0 in the asynchronous correlation spectrum (compare case (a) in the bottom right panel in Figure \ref{fig:correxam}). They indicate how strong the spectral intensity changes at a given spectral position $\nu_1$. A strong synchronous auto peak signals characteristic high changes at the associated spectral position.\\
Often the information gathered from auto peaks is easy to interpret, but not too helpful when compared to the information gained from cross peaks. Cross peaks are not located at the diagonal and show the correlation between changes at two different spectral values $\nu_1$ and $\nu_2$ (case (b) and case (c) in Figure \ref{fig:correxam}). Following \cite{Noda.1990, Noda.2006} there are five Noda rules, which can be used to interpret synchronous and asynchronous 2D correlation cross peaks:

\begin{enumerate}
\item If the sign of a cross peak at a spectral coordinate pair ($\nu_1$, $\nu_2$) of a synchronous 2D correlation spectrum is positive, \textit{i.e.}, $\Phi(\nu_1, \nu_2) > 0$, the spectral intensities measured at $\nu_1$ and $\nu_2$ are changing in the same direction, \textit{i.e.}, both intensities are either increasing or decreasing simultaneously.
\item On the other hand, if the sign of a synchronous cross peak is negative, \textit{i.e.}, $\Phi(\nu_1, \nu_2) < 0$, the spectral intensities measured at $\nu_1$ and $\nu_2$ are changing in the different directions, \textit{i.e.}, one is increasing while the other is decreasing.
\item If the sign of a cross peak at a spectral coordinate pair ($\nu_1$, $\nu_2$) of an asynchronous 2D correlation spectrum is positive, \textit{i.e.}, $\Psi(\nu_1, \nu_2) > 0$, the spectral intensity measured at $\nu_1$ varies before that measured at $\nu_2$ with respect to the perturbation.
\item If the sign of an asynchronous cross peak is negative, \textit{i.e.}, $\Psi(\nu_1, \nu_2) < 0$, the spectral intensity measured at $\nu_1$ varies after that measured at $\nu_2$ with respect to the perturbation.
\item However, if the sign of a synchronous cross peak located at the same spectral coordinate ($\nu_1$, $\nu_2$) is negative, \textit{i.e.}, $\Phi(\nu_1, \nu_2) < 0$, the above two rules are reversed.
\end{enumerate}

As an starting example we have a look at the synchronous 2D correlation spectrum of the FuranMale dataset (Figure~\ref{fig:asynbasic}): We see three (positive) auto peaks at (1575 cm$^{-1}$, 1575 cm$^{-1}$), (1585 cm$^{-1}$, 1585 cm$^{-1}$) and (1600 cm$^{-1}$, 1600 cm$^{-1}$) as well as two negative cross peaks at (1575 cm$^{-1}$, 1585 cm$^{-1}$) and (1575 cm$^{-1}$, 1600 cm$^{-1}$) as well as one positive crosspeak at (1585 cm$^{-1}$, 1600 cm$^{-1}$). The other three cross peaks are redundant, because a synchronous homo 2D correlation spectrum is always symmetric with respect to the diagonal.  The auto peaks indicate that all three Raman bands are changing during the heating. The positive cross peak at (1585 cm$^{-1}$, 1600 cm$^{-1}$) tells us that the bands at 1585 cm$^{-1}$ and 1600 cm$^{-1}$ are changing in the same direction (Noda rule 1), while the two negative cross peaks at (1575 cm$^{-1}$, 1585 cm$^{-1}$) and (1575 cm$^{-1}$, 1600 cm$^{-1}$) unveil that the band at 1575 cm$^{-1}$ is changing in a different direction when compared to the band at 1585 cm$^{-1}$ and 1600 cm$^{-1}$ (Noda rule 2). If this information about the three bands is combined with the information from the one-dimensional Raman spectrum that the band at 1585 cm$^{-1}$ is increasing in intensity during the heating, it becomes clear that the band at 1600 cm$^{-1}$ is also increasing in intensity while the band at 1575  cm$^{-1}$ is falling in intensity during the heating.\\
For further conclusions from the Raman spectra the reader is referred to \cite{Geitner.2015}. Further examples and extended discussion on the interpretation of 2D correlation spectra can be found in \cite{Noda.1993, Czarnecki.1998, Noda.2006, Noda.2012, Noda.2014}.\\

\section{Technical aspects}

\subsection{Software versions and hardware setup}
\proglang{R} as a software language and its software packages are being actively developed. Thus, this manuscript can only be a snapshot regarding the ongoing development process. For information on the latest version of \pkg{corr2D} the user is referred to the online documentation of \pkg{corr2D} \citep{Geitner.2016b}.\\
The current \proglang{R} version is 3.4.1, the version of the \proglang{R} core packages \pkg{parallel}, \pkg{stats}, \pkg{graphics} and \pkg{grDevices} are also 3.4.1 \citep{RCoreteam.2016}, the version of \pkg{doParallel} is 1.0.10 \citep{RevolutionAnalytics.2015b}, the version of \pkg{foreach} is 1.4.4 \citep{RevolutionAnalytics.2015}, the version of \pkg{fields} is 9.0 \citep{Nychka.2016}, the version of \pkg{mmand} is 1.5.1 \citep{Clayden.2016}, the version of \pkg{rgl} is 0.98.1 \citep{Adler.2016} and the version of \pkg{colorspace} is 1.3.2 \citep{Ihaka.2016}. The version of \pkg{corr2D} discussed here is 0.1.12.\\
The calculation speed test of \code{corr2d()} depending on the input matrix dimensions and the number of processor cores used was done using the function \code{microbenchmark()} from package \pkg{microbenchmark} \citep{Mersmann.2015} with 10 calculation cycles. The test system was a 64-bit Windows 7 (SP1) setup with 8 GB of random access memory (RAM) and a quadcore Intel Core i5-2450M processor.\\
The profiling of the two functions \code{corr2d()} and \code{plot_corr2d()} from \pkg{corr2D} was done using a 64-bit Windows 7 (SP1) system with Intel Core2 Duo CPU E7400 @ 2.80 GHz, 4 GB of RAM as well as an Intel G35 Express Chipset as graphical processing unit and the package \pkg{profr} \citep{Wickham.2014}. The 2D correlation software 2DShige was tested with the same system.\\

\FloatBarrier
\subsection[corr2d(): From equations to R~code]{\code{corr2d()}: From equations to \proglang{R}~code}
\label{sec:eqtoRcode}
In the following two sections we will have a closer look at the correlation function \code{corr2d()}, how it handles the input perturbation variable, how it generates the dynamic spectra, how it applies scaling techniques to 2D correlation spectroscopy, how the parallel correlation process works and how the \proglang{R}-script compares to other 2D correlation software in terms of speed and user-friendliness.\\
As described in the introduction it is important that the discrete perturbation values $T$ are equidistant. This requirement is often difficult to fulfill when working with real world data, especially for big datasets with a lot of perturbation values or perturbation values which are hard to adjust like the pH value. To overcome this problem there are three approaches:

\begin{enumerate}
\item Ignore the requirement for equidistant perturbation values and use the data as it is.
\item Use modified correlation equations as described in \cite{Noda.2003} to account for the uneven sampling of the perturbation variable. 
\item Interpolate the perturbation values and the associated spectral data to get an equidistant distribution of the perturbation values. 
\end{enumerate}

The first approach is the usual go-to solution, because it takes no complex interpolation and yields reasonable results if the perturbation value distribution is nearly equidistant. The approach fails if the perturbation values are distributed very unevenly over a large observation window. In this case it becomes necessary to interpolate the perturbation values to get an equidistant distribution to use the correlation equations for evenly sampled perturbation values (Equations \ref{eq:int2d}, \ref{eq:HTsyn} and \ref{eq:HTasyn}).\\
For the interpolation \code{corr2d()} can use a wide variety of interpolation algorithms, which can be specified at the \code{Int} argument. In a simple scenario this could be a linear function modeled by \code{approxfun()}. The default interpolation function for \code{corr2d()} is the cubic Hermite function \code{splinefun()} from package \pkg{stats}. The interpolation process consists of three steps: An interpolation to get $m$ (specified by argument \code{N}) equidistant perturbation values, a value wise interpolation of the spectral dataset using the interpolation function given by \code{Int} and the calculation of the new spectral dataset using the interpolated spectral dataset and the interpolated perturbation values.

\begin{Schunk}
\begin{Sinput}
R> TIME <- seq(min(Time), max(Time), length.out = N)
R> tmp <- apply(Mat1, 2, function(y) Int(x = Time, y = y))
R> Mat1 <- sapply(tmp, function(x) x(TIME))
R> Time <- TIME
\end{Sinput}
\end{Schunk}

The code snippet shows the interpolation process. The old minimum and maximum perturbation values from \code{Time} are used to generate the new perturbation values \code{TIME}. The spectral dataset in \code{Mat1} is interpolated column wise (which is equal to spectral value wise) using the old perturbation values from \code{Time} and interpolation specified by \code{Int} to get a list of functions \code{tmp} describing the behavior of the spectral intensity at every spectral position along the perturbation axis. This function is then used together with the new perturbation values \code{TIME} to calculate the interpolated spectral dataset. In addition the new equidistant perturbation values get saved.\\
This approach to get an evenly sampled perturbation variable is very flexible and allows the use of a wide variety of interpolation functions. This flexibility can be used to get good results from an otherwise sub optimal dataset.\\

The next step on the way to calculate 2D correlation spectra is the calculation of the dynamic spectra. The dynamic spectra are built from the original dataset by subtracting a reference spectrum (see Equation~\ref{eq:ref}). Thus, the dynamic spectra show the changes with respect to the chosen reference spectrum.\\
The reference spectrum can be any spectrum as long as it has the same number of spectral values as the input data. When doing a 2D correlation analysis on a new dataset usually the perturbation mean spectrum is chosen as the reference spectrum. Other reference spectra could be the starting or the end spectrum as well as an external spectrum which is not part of the original dataset. The choice of the reference spectrum depends on the analytical problem and what spectral changes should be highlighted. The correlation function \code{corr2d()} takes one vector containing the reference spectrum at argument \code{Ref1} (and a second reference spectrum at \code{Ref2} when doing a hetero correlation analysis) and builds the dynamic spectra through subtracting via the \code{sweep()} function following Equation~\ref{eq:dynspec}. If no reference spectrum is specified the perturbation mean spectrum is calculated from the input matrix \code{Mat1} (or \code{Mat2}) according to Equation \ref{eq:ref} and subtracted instead. In this case the dynamic spectra become the mean-centered spectra.

\begin{Schunk}
\begin{Sinput}
R> if (is.null(Ref1)) {
+   Ref1 <- colMeans(Mat1)
+   Ref2 <- colMeans(Mat2)
+ }
R> Mat1 <- sweep(Mat1, 2, Ref1, "-")
R> if (Het == FALSE) {
+   Mat2 <- Mat1
+ } else {
+   Mat2 <- sweep(Mat2, 2, Ref2, "-")
+ }
\end{Sinput}
\end{Schunk}

The final step before doing the actual correlation is applying scaling techniques. Unfortunately, Equations~\ref{eq:scalesyn} and~\ref{eq:scaleasyn} discussed in Section~\ref{sec:theo} are not ideal to be directly incorporated into \proglang{R}~code. The number of calculations necessary to scale a synchronous (or asynchronous) homo correlation spectrum of dimensions [$n \times n$] is $n^2$, where $n$ is the number of spectral values. Because homo correlation spectra are symmetric regarding the diagonal the number of calculations necessary reduces to $(n^2 + n)/2$. Therefore, the time needed to scale a complex homo correlation spectrum consisting of one synchronous and one asynchronous spectrum scales with $n^2 + n$.

\begin{Schunk}
\begin{Sinput}
R> if (scaling > 0) {
+   sd1 <- apply(Mat1, 1, sd)
+   sd2 <- apply(Mat2, 1, sd)
+   Mat1 <- Mat1 / (sd1^scaling)
+   Mat2 <- Mat2 / (sd2^scaling)
+ }
\end{Sinput}
\end{Schunk}

To circumvent the quadratic time scaling \code{corr2d()} scales the dynamic spectra before doing the correlation. The scaling of mean-centered dynamic spectra by the standard deviation is called auto-scaling. The number of calculations necessary to get auto-scaled dynamic spectra is $m\cdot n$, where $m$ is the number of spectra in the dataset and $n$ is the number of spectral values within each spectrum. Thus, the time needed to auto-scale a dataset is proportional to $m\cdot n$.\\
When comparing the two calculation times it becomes clear that applying the scaling before the correlation is faster than applying the scaling after the correlation as along $m$ is smaller than $n$. In other words as long as the number of spectra within the dataset is smaller than the number of spectral values in each spectrum it is faster to apply the scaling before the correlation. This condition is often met because the usual 2D correlation analysis features 10-30 different perturbations values, while each spectrum consists of hundreds of spectral values \textit{e.g.}, 1024. Therefore, the function \code{corr2d()} applies the scaling before the correlation.\\

The function \code{corr2d()} uses the FT approach described by Equations \ref{eq:FT} and \ref{eq:int2d} to calculate 2D correlation spectra. To do the DFT \code{corr2d()} uses the FFT provided by the function \code{fft()} from package \pkg{stats}. This type of DFT is much faster than normal DFT when the number perturbation values has a lot of factors. Thus, \code{corr2d()} tries to interpolate to 4, 8, 16, $\dots$, $2^n$ perturbation values when interpolating the perturbation axis.\\
A problem when implementing the FFT in a linear \code{for} loop is the speed of the calculations, which drops dramatically when processing a dataset with a large number of spectral values. Therefore, we decided to parallelize the FFT calculations using the \pkg{foreach} package. Parallel processing using the multicore structure of modern computers can lead to significant reduced calculation times when doing a large number of similar operations.

\begin{Schunk}
\begin{Sinput}
R> ft1 <- foreach::foreach(i = 1:NCOL(Mat1), .combine = 'cbind') %dopar% {
+   fft(Mat1[, i])[1:(NROW(Mat1) - 1) %/% 2 + 1]
+ }
R> if (Het == FALSE) {
+   ft2 <- ft1
+ } else {
+   ft2 <- foreach::foreach(i = 1:NCOL(Mat2), .combine = 'cbind') %dopar% {
+     fft(Mat2[, i])[1:(NROW(Mat2) - 1) %/% 2 + 1]
+   }
+ }
\end{Sinput}
\end{Schunk}

The code snippet highlights the FFT implementation in \code{corr2d()}. With the functions \code{foreach()} and \code{\%dopar\%} the calculation process is split up according to the number of available cores so that the \code{fft()} calculations will be done in parallel. The result of the parallel calculations is a matrix containing the FTs for every spectral value over the whole perturbation range. This matrix is the discrete analogue to $\widetilde{Y}(\nu, \omega)$ from Equation~\ref{eq:FT}. If a homo correlation is done the matrix is duplicated, for a hetero correlation the FFT is also done for the second matrix of dynamic spectra.\\
\code{corr2d()} discards half of the calculated FT values. This step can be explained by the symmetry of the FT of real input data. If a signal $y$ gets Fourier transformed, the result is a complex signal $\mathcal{F}(y)$ consisting of a real and an imaginary part. If the input signal $y$ is real, the real part of the FT is always even, while the imaginary part is always odd. In addition the FT values feature Hermitian symmetry $\mathcal{F}(y)(\omega) = \mathcal{F}^*(y)(m-\omega)$ as seen in Equation~\ref{eq:symmeFT}. Following this symmetry condition only half the values of a FT of a real signal are needed to fully describe the information in the corresponding FT. Thus, it is reasonable to only use one half of the FT values when doing a correlation analysis of real data. To extract the correct amplitudes of the Fourier frequencies, the corresponding amplitudes have to be doubled when utilizing the Hermitian symmetry to simplify the calculation process. To account for this, the function \code{corr2d()} uses all FT values were $\omega \neq 0$ twice.

\begin{Schunk}
\begin{Sinput}
R> cl <- parallel::makeCluster(corenumber)
R> doParallel::registerDoParallel(cl)
R> FT <- matrix(Norm * parallel::parCapply(cl, ft1, get("%*%"), Conj(ft2)),
+   NCOL(ft1), NCOL(ft2), byrow = TRUE)
\end{Sinput}
\end{Schunk}

The correlation procedure follows Equation~\ref{eq:int2d} and is parallelized using the function \code{parCapply()} from package \pkg{parallel}. \code{parCapply()} directs the column wise matrix multiplication to the different cores which are registered before the calculation. The correlation is done using the Fourier transformed dynamic spectra for every spectral value pair $(\nu_1, \nu_2)$. In addition the resulting correlation matrix gets normalized by factor \code{Norm} which gets specified in the input to \code{corr2d()}. The default normalization is the factor $1 / (\pi \cdot (m - 1))$ where $m$ is the number of sampled perturbation values. The real part of the complex output matrix \code{FT} is the synchronous correlation spectrum $\Phi(\nu_1, \nu_2)$ while the imaginary part is the asynchronous correlation spectrum $\Psi(\nu_1, \nu_2)$ (see Equation~\ref{eq:int2d}).\\

\FloatBarrier
\subsection{Speed test of calculation}

The parallelization speeds up the calculation process. To measure the influence of the parallelization and the influence of the input matrix dimensions we designed a small speed test. The speed of \code{corr2d()} was measured using \code{microbenchmark()} from package \pkg{microbenchmark} with 10 calculation cycles. The input matrix was simulated by \code{sim2ddata()} (from \pkg{corr2D}) with 200, 400, 600, 1000, 4000 and 8000 spectral values $n$ and 5, 10, 20, 100 and 500 perturbation values $m$. The simulated spectral data of the consecutive first order reaction was also used in \cite{Noda.2014}. The calculations were done with 1, 2 or 4 cores, respectively. The calculation and plotting time used by 2DShige was estimated 10 times using the Windows task manager.

\begin{table}[tb]
\centering
\begin{tabular}{|rcr|rrrrrr|p{0.4cm}|}
  \hline
  & & &\multicolumn{6}{c|}{$n$} & \\
 & & $m$ & 200 & 400 & 600 & 1000 & 4000 & 8000 & \\
 \hline
 & \parbox[c]{1mm}{\multirow{5}{*}{\rotatebox[origin=c]{90}{1 core}}} & 5 & 0.36 & 0.50 & 0.65 & 1.05 & 5.55 & 16.37 & \parbox[c]{1mm}{\multirow{5}{*}{\rotatebox[origin=c]{90}{in s}}} \\ 
   &  & 10 & 0.36 & 0.50 & 0.66 & 1.10 & 5.76 & 18.21 &  \\ 
  (a) &  & 20 & 0.36 & 0.50 & 0.67 & 1.11 & 6.17 & 18.61 &  \\ 
   &  & 100 & 0.63 & 0.98 & 0.92 & 1.33 & 9.74 & 33.04 &  \\ 
   &  & 500 & 0.48 & 0.82 & 1.23 & 2.57 & 26.53 & 105.23 &  \\ 
   \hline
\hline
 & \parbox[c]{1mm}{\multirow{5}{*}{\rotatebox[origin=c]{90}{2 cores}}} & 5 & 0.56 & 0.69 & 0.80 & 1.18 & 5.03 & 15.09 & \parbox[c]{1mm}{\multirow{5}{*}{\rotatebox[origin=c]{90}{in s}}} \\ 
   &  & 10 & 0.62 & 0.69 & 0.83 & 1.18 & 5.16 & 15.88 &  \\ 
  (b) &  & 20 & 0.56 & 0.68 & 0.82 & 1.21 & 5.40 & 16.87 &  \\ 
   &  & 100 & 0.59 & 0.73 & 0.89 & 1.33 & 7.50 & 27.01 &  \\ 
   &  & 500 & 0.66 & 0.92 & 1.27 & 2.15 & 18.46 & 74.43 &  \\ 
   \hline
\hline
 & \parbox[c]{1mm}{\multirow{5}{*}{\rotatebox[origin=c]{90}{4 cores}}} & 5 & 1.00 & 1.11 & 1.21 & 1.62 & 5.45 & 15.49 & \parbox[c]{1mm}{\multirow{5}{*}{\rotatebox[origin=c]{90}{in s}}} \\ 
   &  & 10 & 0.99 & 1.09 & 1.22 & 1.59 & 5.55 & 16.11 &  \\ 
  (c) &  & 20 & 0.99 & 1.09 & 1.23 & 1.63 & 5.88 & 17.20 &  \\ 
   &  & 100 & 1.02 & 1.13 & 1.31 & 1.73 & 7.82 & 26.81 &  \\ 
   &  & 500 & 1.10 & 1.42 & 1.70 & 2.62 & 18.33 & 72.34 &  \\ 
   \hline
\hline
 & \parbox[c]{1mm}{\multirow{5}{*}{\rotatebox[origin=c]{90}{ratio$_{4/1}$}}} & 5 & 2.81 & 2.21 & 1.85 & 1.54 & 0.98 & 0.95 & \parbox[c]{1mm}{\multirow{5}{*}{\rotatebox[origin=c]{90}{/}}} \\ 
   &  & 10 & 2.75 & 2.16 & 1.84 & 1.45 & 0.96 & 0.89 &  \\ 
  (d) &  & 20 & 2.75 & 2.16 & 1.83 & 1.46 & 0.95 & 0.92 &  \\ 
   &  & 100 & 1.62 & 1.15 & 1.42 & 1.30 & 0.80 & 0.81 &  \\ 
   &  & 500 & 2.30 & 1.73 & 1.39 & 1.02 & 0.69 & 0.69 &  \\ 
   \hline
\end{tabular}
\caption{Mean calculation times (in seconds) needed by \code{corr2d()} to do a homo 2D correlation analysis using 1 (a), 2 (b) or 4 (c) processor cores and a ratio (d) calculated from the mean calculation times using 4 and 1 processor cores. The table includes the number of perturbation values $m$ by rows and the number of spectral values $n$ by columns.} 
\label{tab:core2}
\end{table}
The result of the speed test can be seen in Table~\ref{tab:core2}(a)-(c). As one can see from the table the parallelization speeds up the calculation of large 2D correlation spectra while it slows down the calculation for smaller 2D spectra. Table~\ref{tab:core2}(d) illustrates this observation. This observation can be explained by how parallel computing works: For a parallelized calculation every core calculates only a small part of the problem. These small tasks need to be transferred to the calculating cores and later the results of all these small calculations need to be put together to get the result for the parallelized calculation. The time needed by this processes increases the more the original tasks is split up. Thus, a parallel computation is always a tradeoff between speeding up the calculation process and increasing the amount of traffic needed to organize the parallel computation. This tradeoff can be seen at the parallel calculations done by \code{corr2d()}: \code{corr2d()} needs more time calculating the correlation in parallel when the input matrix is small compared to a serialized calculation with only one processor core. For larger input matrices the effect tips and \code{corr2d()} is faster using more cores. For small input matrices the correlation speed differences are hardly noticeable (0.8 s vs 1.4 s for a [500 $\times$ 400] matrix) while the speed differences are much more important if the input matrices get bigger and the calculation times get longer (26.5 s vs 18.3 s for a [500 $\times$ 4000] matrix).\\
The number of computation steps necessary also depends on how the input for the correlation integral is calculated. The number of computation steps varies for the FT and the HT approach. For $m$ discrete input spectra the FFT needs $4 \cdot m \cdot log_2(m)$ steps (using Cooley-Turkey algorithm \citep{Cooley.1965}) while the HT needs $m^2$ steps. Therefore, the HT is faster for smaller datasets while the FFT is faster for larger datasets. Because calculation speed matters more for larger datasets, \code{corr2d()} uses the FT approach. For details on the comparison between the FT and the HT approach the reader is referred to \cite{Noda.2000b}.\\
The speed of the calculation drops off significantly if the RAM limit dedicated to \proglang{R} is reached. In this case \code{corr2d()} needs to save results outside the RAM which takes a large amount of time when compared to saving data inside the RAM. To overcome this problem more RAM needs to be dedicated to \proglang{R} by using \code{memory.limit()}.

\begin{Schunk}
\begin{Sinput}
R> simdata <- sim2ddata(4000, seq(0, 10, length.out = 100))
R> library("profr")
R> prof2d <- profr(c(speedtwod <- corr2d(simdata, Time =
+   as.numeric(rownames(simdata)), scaling = 0.5), plot_corr2d(speedtwod)),
+   interval = 0.005)
\end{Sinput}
\end{Schunk}

\begin{figure}[tp]
\centering
\includegraphics[scale = 0.9]{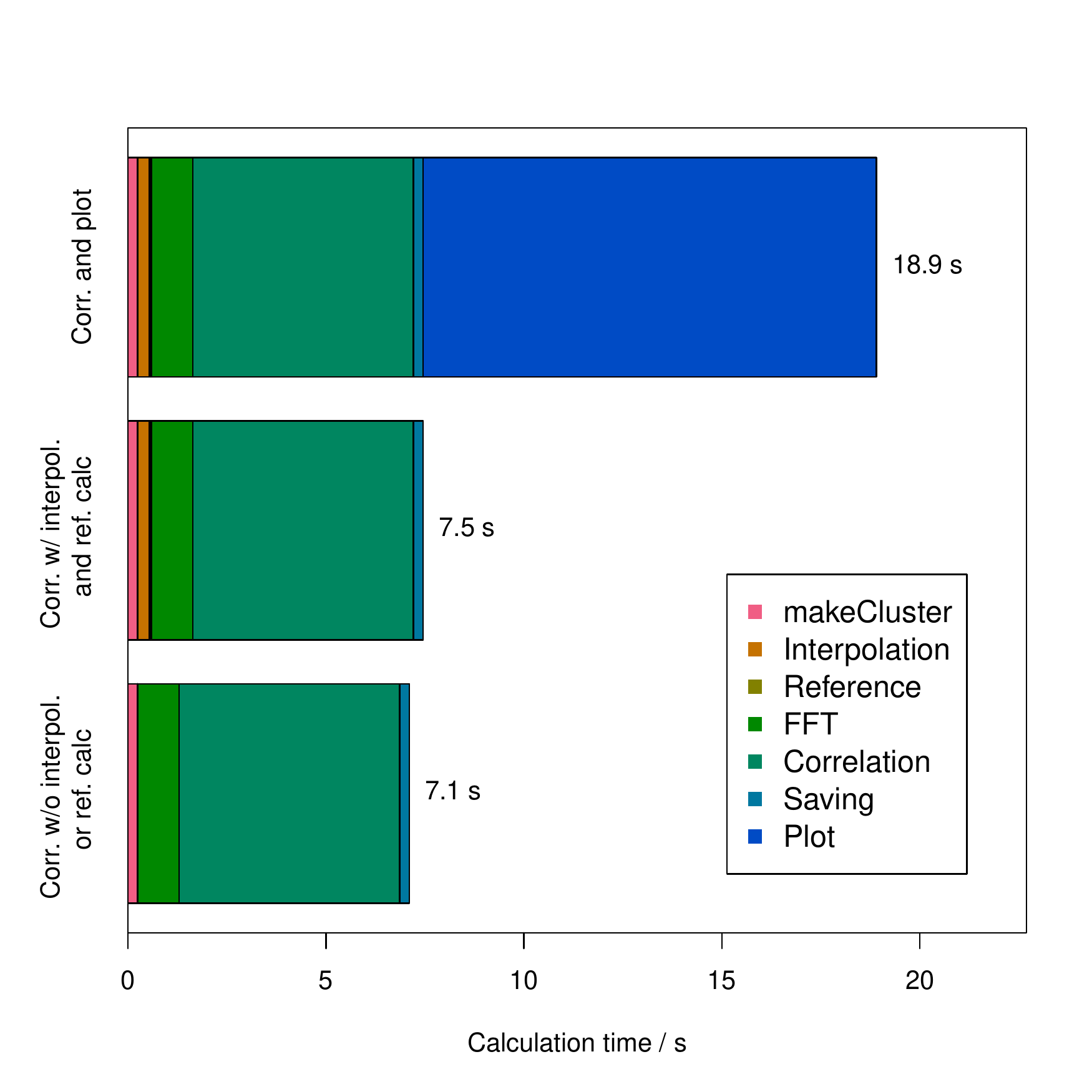}
\caption{Calculation time used by \code{corr2d()} to calculate a [4000 $\times$ 4000] complex correlation spectrum from a [100 $\times$ 4000] input matrix. The colored bars show how much calculation time is used by each single calculation step.}
\label{fig:speedcomp}
\end{figure}

\begin{figure}[tp]
\centering
\includegraphics[scale = 0.9]{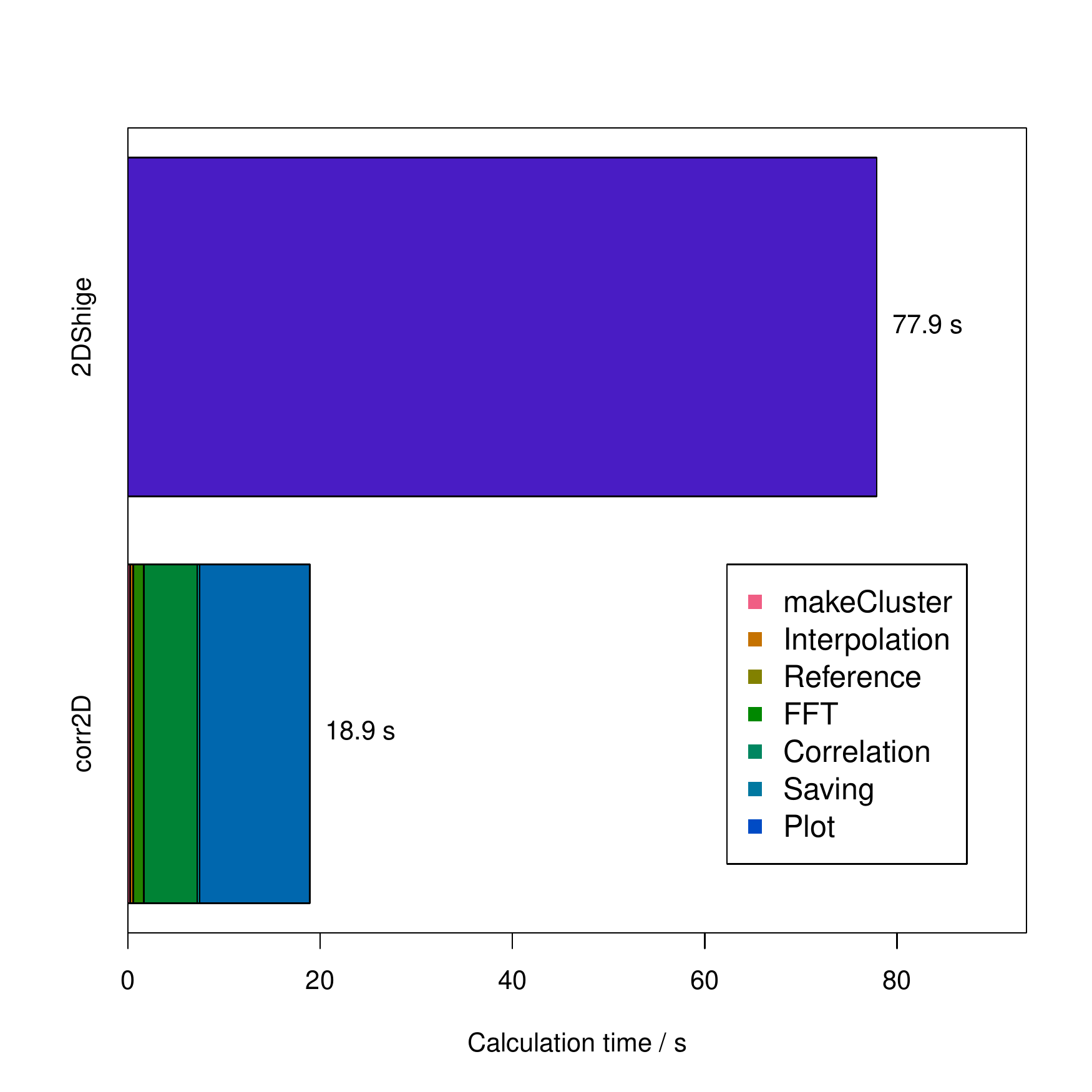}
\caption{Comparison between \pkg{corr2D} and 2DShige. Calculation and plotting time used to evaluate a [4000 $\times$ 4000] complex correlation spectrum from a [100 $\times$ 4000] input matrix. The colored bars show how much calculation time is used by each single calculation step.}
\label{fig:speedcomp2}
\end{figure}

To illustrate how much time is spend in each step, the correlation function \code{corr2d()} and plotting function \code{plot\_corr2d()} were profiled using \code{profr()} from package \pkg{profr}. The results can be seen in Figure~\ref{fig:speedcomp}. Most of the calculation time used by \code{corr2d()} is needed for the parallel matrix multiplication (5.6 s) followed by the FFT (1.1 s). The interpolation process to get 128 perturbation values (0.3 s) and the application of scaling techniques (0.05 s) are faster. The plotting of the resulting synchronous 2D correlation spectrum also needs time (11.4 s) because the complete spectrum has 16 MPixel.\\
To evaluate the script in a larger scheme we tried to compare the \proglang{R}~package \pkg{corr2D} to the software 2DShige \citep{Morita.2005} which is available free of charge from the internet. Because 2DShige is a stand-alone program the timing of its calculations are not as precise as the timings from the profiling of \code{corr2d()} and \code{plot\_corr2d()}. The calculation and plotting time used by 2DShige was estimated 10 times using the Windows task manager. The input matrix was the same simulated [100 $\times$ 4000] data matrix used for the profiling.\\
2DShige took 77.9 s to calculate the [4000 $\times$ 4000] complex 2D correlation spectrum and plot the synchronous 2D correlation spectrum. Unfortunately, it is not possible to give calculation times for every calculation step. Comparing the calculation times need by 2DShige and \pkg{corr2D} it becomes clear that the \proglang{R}~package is faster when calculating a large 2D correlation spectrum. Figure~\ref{fig:speedcomp2} illustrates the results.\\
Overall, the profiling of \code{corr2d()} and \code{plot\_corr2d()} and the speed test done for \code{corr2d()} prove that it is useful to parallelize the FFT and the matrix multiplication. Other calculation steps (interpolation, reference spectrum calculation and scaling) are much faster than the FFT and the matrix multiplication and thus do not need a parallelization. The parallel calculation approach is faster when compared to a serialized approach for large input matrices. The \proglang{R}~code also compares favorably to other 2D correlation software in terms of speed and transparency.\\

\subsection[Visualizing 2D correlation spectra]{Visualizing 2D correlation spectra}

The following section describes the programming and design details considered for the two plotting function \code{plot\_corr2d()} and \code{plot\_corr2din3d()}. In general the representation of 2D correlation spectra is designed with 2D NMR plots in mind. 2D NMR plots also show 3D data with two frequency axes (the so called chemical shift $\delta$) and contour levels to represent crosspeaks. 2D NMR spectroscopy is a common technique in chemistry labs. Thus, the plotting function \code{plot\_corr2d()} tries to mimic the appearance of 2D NMR plots to make the data accessible for beginners in the field of 2D correlation spectroscopy.\\
Most of the arguments in \code{plot\_corr2d()} are inspired by the arguments in \code{plot()} or \code{image()} to allow users already experienced with \proglang{R} a smooth transition. The function \code{plot\_corr2d()} takes an object \code{Obj} of class \code{corr2d} as produced by the correlation function \code{corr2d()} and extracts the different parameters needed for depicting the 2D correlation spectra contained in the given object.\\
Firstly, the function \code{plot\_corr2d()} saves the prior user defined graphics parameters to restore them at the end. Secondly, \code{plot\_corr2d()} also extracts the graphical parameters asigned by the user under the \code{...} argument. This procedure enables the function to later use these parameters in different plotting functions to adjust the appaerance of the entire 2D correlation plot.

\begin{Schunk}
\begin{Sinput}
R> par_old <- par(no.readonly = TRUE)
R> getparm <- list(...)
R> graphparm <- utils::modifyList(par(), getparm)
\end{Sinput}
\end{Schunk}

After the extraction of the graphical parameters, the 2D plot function subsets the spectral variable axes \code{Obj$Wave1} (\textit{x}-axis) and \code{Obj$Wave2} (\textit{y}-axis) to the window range defined by the arguments \code{xlim} and \code{ylim}.

\begin{Schunk}
\begin{Sinput}
R> if (is.null(xlim)) {
+   Which1 <- 1:NROW(what)
+ } else {
+   Which1 <- which(xlim[1] < Obj$Wave1 & Obj$Wave1 < xlim[2])
+ }
R> if (is.null(ylim)) {
+   Which2 <- 1:NCOL(what)
+ } else {
+   Which2 <- which(ylim[1] < Obj$Wave2 & Obj$Wave2 < ylim[2])
+ }
\end{Sinput}
\end{Schunk}

\begin{figure}[tbp]
\centering
\includegraphics{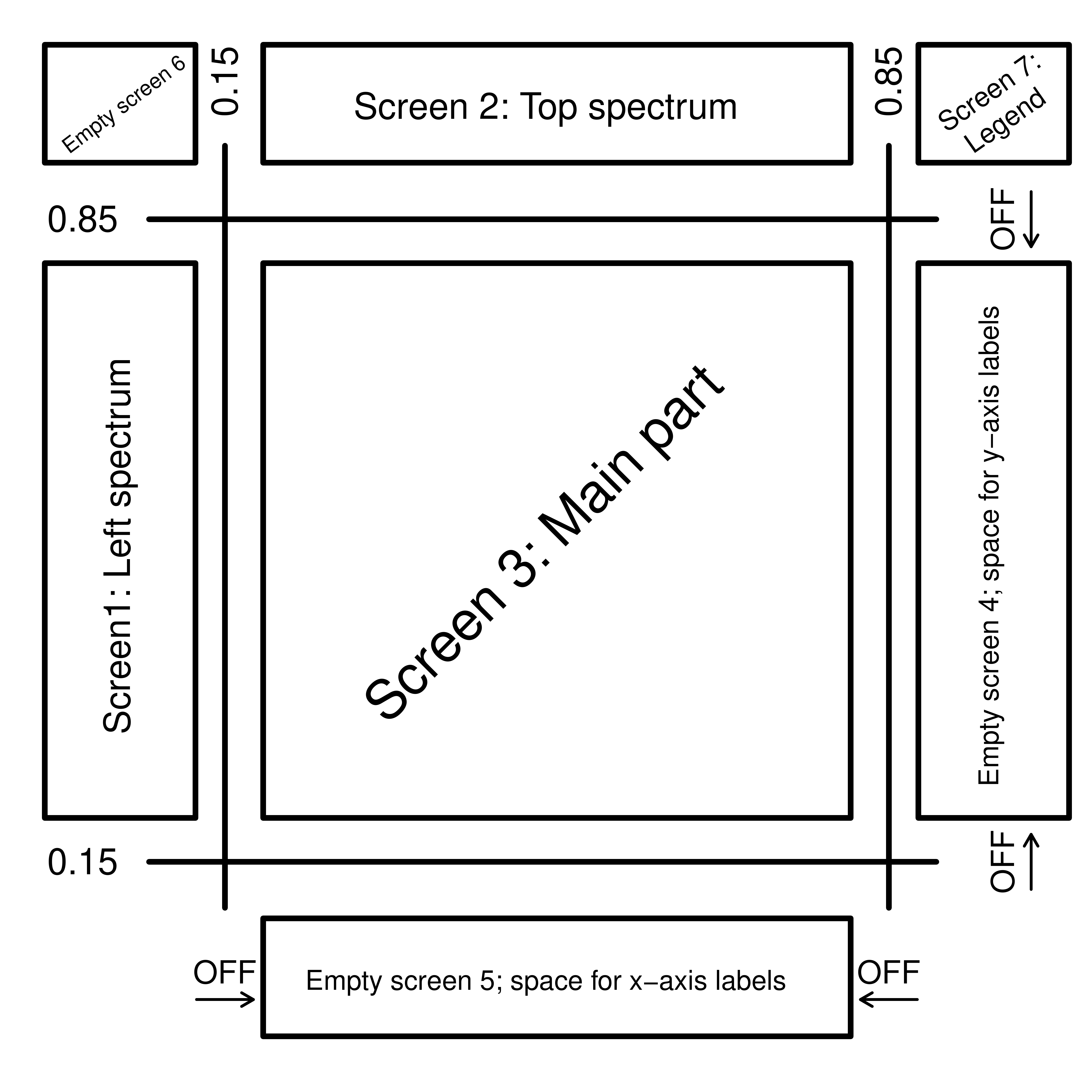}
\caption{Schematic representation of the seven screens generated and used by \code{plot\_corr2d()} to graphical represent 2D correlation spectra. The offset \code{OFF} is used during development to adjust the screen ratios.}
\label{fig:splitscreen}
\end{figure}

The 2D plot function uses the \code{contour()} or \code{image()} function in a \code{split.screen()} environment (all from package \pkg{graphics}) to generate the 2D plot. The plot device is split into seven screens (see Figure~\ref{fig:splitscreen}), which will be filled with two 1D spectra via a simple line plot (screen 1 and 2), the central 2D plot (screen 3), generated by the \code{contour()} or \code{image()} function with \textit{x}- and \textit{y}-axis as well as their labels and the color legend in the top right (screen 7). The remaining three screens remain (up to now) empty. On exit \code{plot\_corr2d()} closes all but screen 3, thus further \code{plots}, \code{lines} and \code{points} can be added to the central plot or data can be read out interactively using the \code{locator()} function. The code snippet shows the creation of the \code{split.screen()} environment. The offset \code{OFF} is used during development to adjust the ratio between the main part (screen 3) and the surrounding screens.

After the creation of the different screens the function \code{plot\_corr2d()} switches to the different screens and fills them with plots, labels and legends. Screen 1 and 2 are used to plot the 1D reference spectra to the top (\code{specx}) and to the left (\code{specy}) of the main part in screen 3. By default the reference spectra \code{Obj$Ref1} and \code{Obj$Ref2} are plotted, but the user can assign any data to be plotted on the axes. As \textit{x}-axis for the two plots the previously subset spectral variable axes \code{Obj$Wave1[Which1]} and \code{Obj$Wave2[Which2]} are used to align the 1D spectra with the central 2D plot. Thus, the reference spectra (specified at \code{specx} and \code{specy}) must be of same length as the spectral variable axes. The reference spectrum on the \textit{y}-axis is rotated by switching the \textit{x}- and \textit{y}-axes around as well as by adjusting the new \textit{x}-values accordingly. The plotting of the reference spectra can be suppressed by setting \code{specx} and/or \code{specy} \code{NULL}. The arguments defined in \code{par()} suppress the plotting of any axes or labels, which are not needed because they are added to the axes of the 2D correlation spectrum. The line width of the 1D plots can be adjusted by asigning the \code{lwd} argument.

\begin{Schunk}
\begin{Sinput}
R> if (!is.null(specy)) {
+   screen(1)
+   par(xaxt = "n", yaxt = "n", mar = c(0, 0, 0, 0), bty = "n", yaxs = "i")
+   plot.default(x = max(specy[Which2]) - specy[Which2],
+     y = Obj$Wave2[Which2], type = "l",
+     lwd = graphparm$lwd + 1, ann = FALSE)
+ }
R> if (!is.null(specx)) {
+   screen(2)
+   par(xaxt = "n", yaxt = "n", mar = c(0, 0, 0, 0), bty = "n", xaxs = "i")
+   plot.default(x = Obj$Wave1[Which1], y = specx[Which1],
+     type = "l", lwd = graphparm$lwd + 1, ann = FALSE)
+ }
\end{Sinput}
\end{Schunk}

The centerpiece of the 2D plot is the 2D correlation spectrum depicted in screen 3. To get a flexible plotting function \code{plot\_corr2d()} features a lot of control arguments for the main part. In line with the underlying functions \code{contour()} and \code{image()} which are used for plotting the 2D spectrum \code{plot\_corr2d()} has a \code{zlim} argument to define the range of the \textit{z}-axis. Next the function \code{plot\_corr2d()} builds \code{N} contour or image levels, which are evenly distributed along a modified \textit{z}-axis. The levels are calculated using the maximum absolute value among all \textit{z}-values. The color values for each level are calculated using the levels derived from the modified \textit{z}-axis. The 2D plot uses the colors darkblue and cyan for negative \textit{z}-values as well as yellow, red and darkred for positive \textit{z}-values. The colors are taken from function \code{designer.colors()} from package \pkg{fields}. After the calculation of the color values the 'real' contour or image levels are calculated using the borders defined by \code{zlim}.\\
This procedure looks unnecessary complicated at first, but the procedure ensures that the most extreme colors (darkblue for negative values and darkred for positive values) are used for the most extreme absolute \textit{z}-values. The advantage of this approach is that positive and negative levels can be compared with each other by looking at the color code. When using just the number range provided by \code{zlim} this is not possible. As an example consider \textit{z}-values ranging from -1 to 9. These values should be plotted with 9 levels. When using the original \textit{z}-range (-1 to 9) for the calculation of the color code the color darkblue would be assigned to the level at -1 because it is the most extreme negative value and the color darkred would be assigned to the level at 9 because it is the most extreme positive value. By looking at the resulting 2D spectrum one would get the impression that 'the darkblue level is as negative as the darkred level is positive' which is clearly not the case. By using the range defined by the most extreme absolute \textit{z}-value (-9 to 9) for the calculation of the color code the color darkblue gets assigned to the (non-existent) level -9, whereas the color darkred still gets assigned to the level 9. The \textit{z}-value -1 will now be plotted using the color cyan, which indicates a small negative value when compared to the level at 9 which is depicted in darkred.\\
In addition to the already discussed features, \code{plot\_corr2d()} always uses an odd number of contour or image levels. The odd number of contour levels leads to a symmetric distribution of the contour and image levels in an asynchronous homo 2D correlation spectrum. Asynchronous homo correlation spectra are skew-symmetric regarding the diagonal and thus the absolute values of the positive and negative correlation intensities are always identical. The odd number of contour or image levels also accomplishes that positive and negative values are represented by the same number of levels (in reference to the most extreme absolute \textit{z}-value) and that the level around 0 can always be transparent to suppress the plotting of noise. Therefore, it can happen that no blue (or no yellow-red) levels are drawn, because the positive (or negative) values are much larger than the negative (or positive) values that the equal spacing of the levels leads to an agglomeration of all negative (or positive) values inside the level around 0, which will be transparent. To circumvent this problem the number of contour levels has to be increased.\\
The argument \code{Cutout} can be used to set more levels than the central level around 0 transparent. This argument can be used to simplify 2D correlation spectra when starting the interpretation of a new 2D correlation spectrum where one would like to have a look at the strongest correlation first and then gradually work through smaller correlations. Care should be taken when using the \code{Cutout} argument as it can also be used to erase unwanted signals and thus create unrealistic 2D correlation spectra. The code snippet illustrates the how the arguments discussed are implemented into \proglang{R}~code.

\begin{Schunk}
\begin{Sinput}
R> screen(3)
R> if (is.null(zlim)) {
+   zlim <- range(what[Which1, Which2])
+ }
R> if (N%%2 == 0){
+   N <- N + 1
+ }
R> Where <- seq(-max(abs(zlim)), max(abs(zlim)), length.out = N) 
R> if (is.null(Cutout)) {
+   OM <- which(Where < 0)
+   OP <- which(Where > 0)
+ } else {
+   OM <- which(Where <= Cutout[1])
+   OP <- which(Where >= Cutout[2])
+ }
R> COL <- rep("transparent", length(Where))
R> COL[OM] <- fields::designer.colors(col = c("darkblue", "cyan"),
+   n = length(OM))
R> COL[OP] <- fields::designer.colors(col = c("yellow", "red", "darkred"),
+   n = length(OP))
R> COL[(N + 1)/2] <- "transparent"
R> COL <- COL[which(zlim[1] < Where & Where < zlim[2])]
R> Where <- seq(zlim[1], zlim[2], length.out = length(COL))
\end{Sinput}
\end{Schunk}

After the calculation of the color code and the contour or image levels the actual 2D correlation spectrum is drawn either by \code{contour()} or by \code{image()} (both from package \pkg{graphics}). The preferred function can be selected from the logical argument \code{Contour} in \code{plot\_corr2d()}. Both functions use the subset matrix \code{what[Which1, Which2]} as \textit{z}-values and the subset vectors \code{Obj$Wave1[Which1]} and \code{Obj$Wave2[Which2]} as respective \textit{x}- or \textit{y}-values. By default \code{plot\_corr2d()} plots the synchronous 2D correlation spectrum as specified by \code{Re(Obj$FT)}. The normal axes and axis labels are suppressed to allow for a flexible definition by the user. After the plotting of the 2D plot, a white line gets drawn across the 2D spectrum which highlights the main diagonal in a homo 2D correlation spectrum. Afterwards a box is drawn around the 2D plot, the \textit{x}- and \textit{y}-axis are added and the \textit{x}- and \textit{y}-axis labels get added as specified by arguments \code{xlab} and \code{ylab}. The axes of the 2D plot can be suppressed by using the argument \code{axes} in \code{plot\_corr2d()}. The input arguments \code{lwd} and \code{cex} can be used to adjust the line width of the axes and the surrounding box as well as the size of the axes labels.

\begin{Schunk}
\begin{Sinput}
R> par(xaxt = "n", yaxt = "n", mar = c(0, 0, 0, 0),
+   bty = "n", xaxs = "i", yaxs = "i")
R> if (Contour == TRUE){
+   graphics::contour(x = Obj$Wave1[Which1], y = Obj$Wave2[Which2],
+     z = what[Which1, Which2], col = COL, levels = Where,
+     zlim = zlim, drawlabels = FALSE, ...)
+ } else {
+   graphics::image(x = Obj$Wave1[Which1], y = Obj$Wave2[Which2],
+     z = what[Which1, Which2], col = COL, xlab = "",
+     ylab = "", zlim = zlim, ...)
+ }
R> abline(a = 0, b = 1, col = rgb(red = 1, green = 1, blue = 1,
+   alpha = 0.5), lwd = graphparm$lwd)
R> par(xpd = NA, xaxt = "s", yaxt = "s", xaxs = "i", yaxs = "i",
+   cex = graphparm$cex, mar=c(0, 0, 0, 0))
R> box(which = "figure", lwd = graphparm$lwd)
R> if ((axes == 1) | (axes == 3)){
+   axis(side = 1, lwd = graphparm$lwd)
+ }
R> if ((axes == 2) | (axes == 3)){
+   axis(side = 4, las = 2, lwd = graphparm$lwd)
+ }
R> mtext(side = 1, xlab, line = 3.5, cex = graphparm$cex * 1.3,
+   lwd = graphparm$lwd)
R> mtext(side = 4, ylab, line = 3.5, cex = graphparm$cex * 1.3,
+   lwd = graphparm$lwd)
\end{Sinput}
\end{Schunk}

The color code legend is plotted in screen 7. For the legend the function \code{image.plot()} from package \pkg{fields} is used. Most arguments defined in \code{image.plot()} by \code{plot\_corr2d()} are for setting the margins of the plot and arranging the number legend. The color code legend in \code{plot\_corr2d()} has two number values written next to it specifying the 10 \% and 90 \% quantile of the plotted \textit{z}-values. The legend can be turned off by setting the argument \code{Legend} in \code{plot\_corr2d()} to \code{FALSE}. The specified \code{pin} parameter is a small hack to avoid an error produced by \code{image.plot()} in combination the \code{split.screen()} environment. The argument \code{cex.axis} can be defined at the input to adjust the size of the legend labels. After finishing all plotting tasks \code{plot\_corr2d()} changes back to the main 2D plot in screen 3, closes all but screen 3 and restores the old \code{par} parameters. By keeping screen 3 active the user can add points or lines to the central screen and can read out data interactively by using the function \code{locator()}.

\begin{Schunk}
\begin{Sinput}
R> if(Legend == TRUE){
+ 
+   screen(7)
+ 
+   par(pin = abs(par()$pin))
+ 
+   if (Contour == TRUE){
+     fields::image.plot(z = what[Which1,Which2], legend.only = TRUE,
+       smallplot = c(0.15, 0.3, 0.2, 0.8), col = COL,
+       axis.args = list(at = quantile(Where, prob = c(0.1, 0.9)),
+         labels = format(x = quantile(Where, prob = c(0.1, 0.9)),
+         digit = 2, scientific = TRUE), cex.axis = graphparm$cex.axis),
+       zlim = zlim, graphics.reset = TRUE)
+   } else {
+     fields::image.plot(z = what[Which1, Which2],legend.only = TRUE,
+       smallplot = c(0.15, 0.3, 0.2, 0.8), col = COL,
+       axis.args = list(at = range(what[Which1, Which2]),
+         labels = format(x = range(what[Which1, Which2]),
+         digits = 2, scientific = TRUE), cex.axis = graphparm$cex.axis),
+       graphics.reset = TRUE)
+   }
+             
+ }
R> screen(3, new = FALSE)
R> close.screen(c(1,2,4,5,6,7))
R> on.exit(options(par(par_old)), add = TRUE)
\end{Sinput}
\end{Schunk}

The 3D plotting function \code{plot\_corr2din3d()} works a little bit different than the 2D plotting function \code{plot\_corr2d()} because it is meant for creating impressive 3D figures of 2D correlation spectra and not so much for scientific exact representation of 2D correlation spectra. Thus, \code{plot\_corr2din3d()} takes a matrix \code{Mat} (for example the synchronous 2D correlation spectrum \code{Re(Obj$FT)} from an object of class \code{corr2d}), builds arbitrary \textit{x}- and \textit{y}-axis, plots the 3D surface using the function \code{drape.plot()} from package \pkg{fields} and adds user defined 1D spectra to the \textit{x}- and \textit{y}-axis.\\
The creation of arbitrary \textit{x}- and \textit{y}-axis is simply done by using the number of rows and columns in matrix \code{Mat}. To reduce the computational demand when plotting the 3D surface the function \code{plot\_corr2din3d()} has the argument \code{reduce}. The argument \code{reduce} allows the user to resample the input matrix. The resampling is done using the function \code{resample()} from package \pkg{mmand}. When the matrix is resampled the \textit{x}- and \textit{y}-axis are also resampled to match the new matrix. The code snippet shows the axis generation and the resampling.

\begin{Schunk}
\begin{Sinput}
R> par_old <- par(no.readonly = TRUE)
R> x <- 1:NROW(Mat)
R> y <- 1:NCOL(Mat)
R> if (!is.null(reduce)) {
+   Which.x <- (1:length(x))[which(1:length(x)%%reduce == 0)]
+   Which.y <- (1:length(y))[which(1:length(y)%%reduce == 0)]
+   Mat <- mmand::resample(x = Mat, points =
+     list(x = x[Which.x], y = y[Which.y]), kernel = mmand::boxKernel())
+   x <- x[Which.x]
+   y <- y[Which.y]
+ }
\end{Sinput}
\end{Schunk}

If no color pallette is specified at argument \code{Col} than \code{plot\_corr2din3d()} builds the color specification from the \textit{z}-values inside the matrix \code{Mat}. \code{Breaks} describes the numerical divisions of the color scale, which are used by \code{drape.plot()} for plotting the 3D surface. If no \code{zlim} argument, which describes the \textit{z}-axis of the 3D plot, is specified than the \code{zlim} argument is also build from the input matrix. The HCL (hue-chroma-luminance) color space is superior to the widespread RGB (red-green-blue) color space \citep{Zeileis.2009,Stauffer.2015}. Thus, \code{plot\_corr2din3d()} uses the diverging HCL color palette from package \pkg{colorspace} \citep{Ihaka.2016} as default value.

\begin{Schunk}
\begin{Sinput}
R> N <- length(Col)
R> Zero <- 0
R> Max <- max(Mat)
R> Min <- min(Mat)
R> if (N%%2 == 0){
+   Breaks <- c(seq(Min, Zero, length.out =
+     round(N / 2, 0) + 1), seq(Zero, Max,
+       length.out = round(N / 2, 0) + 1)[2:(round(N / 2, 0) + 1)])
+ } else {
+   Breaks <- c(seq(Min, Zero, length.out =
+     round(N / 2, 0) + 2)[1:(round(N / 2, 0) + 1)], seq(Zero, Max,
+       length.out = round(N / 2, 0) + 2)[2:(round(N / 2, 0) + 2)])
+ }
R> if (is.null(zlim)){
+   zlim <- range(Mat, na.rm = TRUE)
+ }
\end{Sinput}
\end{Schunk}

All previously defined parameters are then fused together inside \code{drape.plot()} which plots the 3D surface for the first time. Arguments specified at \code{...} in \code{plot\_corr2din3d()} are handed over to \code{drape.plot()}. The most important arguments are the two viewing angles \code{theta} (\textit{x}-\textit{y} rotation) and \code{phi} (\textit{z}-rotation) as well as the argument \code{border} which takes a color and adds a grid in that color to the 3D surface.\\
\code{drape.plot()} returns a projection matrix which can be used to add a 2D projection of the 3D surface to the bottom of the plot. Unfortunately, \code{drape.plot()} has to be executed once to get the projection matrix. If the 2D surface is simply added to the 3D plot it may overlap with the 3D surface depending on the viewing angles. To circumvent this problem \code{plot\_corr2din3d()} first executes \code{drape.plot()}, than adds the 2D surface and in the end executes \code{drape.plot()} once more to overlay the 2D projection with the 3D surface. The addition of a 2D projection can be specified at argument \code{projection}. The coordinates for the 2D projection are calculated by the function \code{trans3d()} from \pkg{grDevices} and the 2D plot is drawn by \code{polygon()}.

\begin{Schunk}
\begin{Sinput}
R> if (projection == TRUE){
+   WW <- fields::drape.plot(x = x, y = y, z = Mat, col = Col,
+     breaks = Breaks, zlim = zlim, ...)
+   COL <- fields::drape.color(z = Mat, col = Col,
+     zlim = zlim, breaks = Breaks)$color.index
+   for (i in 2:NROW(Mat)) {
+     for (j in 2:NCOL(Mat)) {
+       Points <- grDevices::trans3d(
+         y = y[c(j - 1, j, j, j - 1, j - 1)],
+         x = x[c(i - 1, i - 1, i, i, i - 1)],
+         z = rep(zlim[1], length(5)), pmat = WW)
+       polygon(Points$x, Points$y,
+         border = NA, col = COL[i - 1, j - 1])
+     }
+   }
+     
+   par(new=T)
+     fields::drape.plot(x = x, y = y, z = Mat, col = Col,
+       breaks = Breaks, zlim = zlim, ...)
+     
+   } else {
+     WW <- fields::drape.plot(x = x, y = y, z = Mat, col = Col,
+       breaks = Breaks, zlim = zlim, ...)
+ }
\end{Sinput}
\end{Schunk}

After the 3D surface and its 2D projection are drawn \code{plot\_corr2din3d()} can add custom spectra to the \textit{x}- and \textit{y}-axis. For this reason a new \textit{x}- and/or \textit{y}-axis gets calculated which length is equal to the length of the spectra \code{specx} or \code{specy}. The \code{x}- and \code{y}-arguments for the function \code{line()} are calculated using the aforementioned projection matrix \code{WW} and once again the function \code{trans3d()}. The scaling factors \code{scalex} and \code{scaley} are real numbers and are used to scale the spectra. Positive values at \code{scalex} and \code{scaley} ensure that the spectra are plotted inside the box of the 3D plot (where the 2D projection is also located) whereas negative values place the spectra on the outside of the box. After all plots are done the function \code{plot\_corr2din3d()} restores the old \code{par} parameters.

\begin{Schunk}
\begin{Sinput}
R> if (!is.null(specx)) {
+   if (is.null(scalex)) {
+     scalex <- 1
+   }
+   X <- seq(min(x), max(x), length.out = length(specx))
+   Points.x <- grDevices::trans3d(x = X, y = min(y) + scalex * specx,
+     z = rep(zlim[1], length(X)), pmat = WW)
+   lines(x = Points.x$x, Points.x$y, lwd = 2)
+ }
R> if (!is.null(specy)) {
+   if (is.null(scaley)) {
+     scaley <- 1
+   }
+   Y <- seq(min(y), max(y), length.out = length(specy))
+   Points.y <- grDevices::trans3d(y = Y, x = max(x) - scaley * specy,
+     z = rep(zlim[1], length(Y)), pmat = WW)
+   lines(x = Points.y$x, Points.y$y, lwd = 2)
+ }
R> on.exit(options(par(par_old)), add = TRUE)
\end{Sinput}
\end{Schunk}

\FloatBarrier
\section{Conclusion and outlook}
In this paper we demonstrated how to use our \proglang{R}~package \pkg{corr2D} to calculate and visualize 2D correlation spectra from artificial and real data. The implementation into \proglang{R} offers the advantage to preprocess, correlate, visualize and further analyze data in one open-source program, which was not easily possible before and should help to make the technique of 2D correlation spectroscopy more accessible. For the calculation of the complex correlation spectra we use a parallel fast Fourier transformation approach, which can be adjusted by the user to best fit their needs. The paper also featured a detailed look on the inner workings of our package to help the user understand what is going on during the calculation.\\
In the future we plan on implementing the Hilbert transformation approach into the package as well as advanced 2D correlation spectroscopy techniques like moving-window 2D correlation spectroscopy, 2D codistribution spectroscopy or double 2D correlation spectroscopy. The implementation of advanced techniques into the package will help to make the package more interesting for seasoned spectroscopists and to make new 2D correlation approaches available to a broad user base. Furthermore we also aim to add a graphical user interface via Shiny \citep{Chang.2017} to the package, which will allow users without any programming experience to use the package as well.\\

\section*{Acknowledgments}
We would like to thank the Deutsche Forschungsgemeinschaft (DFG) for funding the priority program SPP 1568 "Design and Generic Principles of Self-healing Materials" and the project PO563/25-2 therein. We would also like to thank Hadley Wickham for his very good tutorials on how to create \proglang{R}~packages and for his helpful \proglang{R}~package \pkg{devtools} \citep{Wickham.2016}. Additionally we also appreciate the helpful comments of the three editors Achim Zeileis, Bettina Gr\"un and Edzer Pebesma.

%
%
\clearpage

\end{document}